\documentclass[10pt,conference,letterpaper]{IEEEtran}

\usepackage{cite}
\usepackage{amsmath}
\usepackage{amssymb}
\usepackage{amsfonts}
\usepackage{mathrsfs}
\usepackage{mathtools}
\usepackage{algorithm}
\usepackage{algorithmic}
\usepackage{graphicx}
\usepackage{textcomp}
\usepackage{xcolor}
\usepackage{url}
\usepackage{booktabs}
\usepackage{bbm}
\usepackage{bm}
\usepackage{multirow}
\usepackage{xspace}
\usepackage{stfloats}
\usepackage{float}
\usepackage[caption=false,font=footnotesize]{subfig}

\newcommand{\para}[1]{\noindent {\bf #1}}

\newcommand{\system}{\textsc{AquaZC}\xspace}
\newcommand{\waveform}{EZCDM\xspace}

\newcommand{\rev}[1]{{\color{black}#1}}

\newcommand{\ignore}[1]{}

\begin{document}

\title{Achieving Rate-Concurrency Balance for Underwater Concurrent Random Access}

\author{
\IEEEauthorblockN{Enqi Zhang, Yuxuan Guo, Weining Li, Linpeng Chen,
Yuetong Chen,\\ Deqing Wang, Lizhao You and Liqun Fu}
\IEEEauthorblockA{School of Informatics, Xiamen University, Xiamen, China}
\IEEEauthorblockA{Corresponding emails: \{deqing,~lizhaoyou\}@xmu.edu.cn}
}

\maketitle

\begin{abstract}
    Underwater acoustic networks face a fundamental rate--concurrency tradeoff: high-rate waveforms (e.g., OFDM, OTFS) are designed for point-to-point links and rely on orthogonal MAC protocols (e.g., TDMA) to avoid collisions, sacrificing concurrency; conversely, collision-resilient waveforms (e.g., CDMA, ZCMod) support uncoordinated access but are inherently rate-limited by spreading or sparse index modulation.
We present \system, a cross-layer concurrent random-access system that combines two new components: (i) \textbf{EZCDM}, an equidistant ZC division-multiplexing waveform that activates multiple cyclic shifts of a ZC root as parallel sub-channels with a tunable rate--robustness tradeoff, and an intra-symbol differential receiver that eliminates the shared multipath channel response without explicit CIR estimation; and (ii) a \textbf{cross-layer link adaptation (LA) framework} featuring beacon-framed random access, user-specific closed-loop power control, and overlap- and CIR-aware common-MS selection. 
Channel-trace- and signal-trace-driven physical-layer experiments combined with PHY-in-the-loop network simulations demonstrate that \system\ achieves significant BER and throughput gains over conventional waveforms and MAC protocols by converting traditionally destructive collisions into decodable concurrent streams.
\end{abstract}



\section{Introduction} \label{sec:intro}

Emerging underwater operations increasingly involve multiple mobile users, including teams of divers and autonomous underwater vehicles (AUVs) \cite{GrandViewResearch2023,Riccardo2021Essential,Colarossi2024,qiu2023optimal,yang2024nerual,hou2025aquascan,tian2026aquascope,tian2025aquavlm}. Such operations naturally favor an uplink-dominated star topology, in which resource-constrained mobile users report data to a resource-rich central gateway. In practice, the gateway faces both time-varying traffic demands and rapidly changing acoustic channels caused by user mobility and environmental fluctuations. Supporting these deployments therefore requires efficient multi-user access that can adapt to both traffic and channel dynamics while sustaining useful data rates.

Existing underwater acoustic (UWA) MAC protocols rely on either coordinated resource allocation or contention-based access, but neither class efficiently accommodates dynamic workloads. Fixed TDMA or FDMA allocations \cite{kredo2009stump} reserve resources in advance, leaving idle slots for inactive users while preventing active users from exploiting them as traffic demands shift. Contention-based schemes such as S-FAMA~\cite{molins2006slotted}, CSMA/CA~\cite{Hwang2016Throughput}, and reservation-free ALOHA variants \cite{chirdchoo2007aloha,Ahn2011Design,Mandal2013A} avoid preallocation, but suffer from long propagation delays or throughput saturation under overlapping transmissions with conventional single-packet reception. Together, these limitations call for on-demand uplink access without per-packet coordination, supported by a PHY capable of separating asynchronous overlapping transmissions at high data rates.

At the PHY layer, existing UWA waveforms face a fundamental rate--concurrency dilemma. High-rate waveforms such as OFDM~\cite{zhou2014ofdm,chen2022underwater}, OCDM~\cite{ouyang2016orthogonal}, and OTFS~\cite{Feng2021otfs} are designed for point-to-point links, where synchronization and accurate channel estimation are needed but possible. However, for synchronous multi-user access, accurate time synchronization and protocol coordination is needed. Maintaining inter-user orthogonality under asynchronous access requires costly multi-user detection or successive interference cancellation, which is vulnerable to errors in rapidly varying UWA multipath channels. 
Conversely, conventional underwater CDMA accommodates asynchronous users through spreading sequences, but is at the cost of low spectral efficiency and substantial channel-estimation overhead \cite{yang2015spatially}.

Recent Zadoff--Chu (ZC)-based waveforms offer a promising direction for narrowing this gap by using the correlation structure of ZC sequences for payload modulation in terrestrial wireless and UWA systems \cite{zhang2022zcnet,vangelista2023golden,zhang2025high}. Their key idea is to encode payloads through cyclic-shift indices within a ZC root: a transmitter activates one or a small subset of shifts from a predefined set, and the receiver identifies the corresponding correlation-peak locations. In particular, ZCMod~\cite{zhang2025high} activates one cyclic shift per symbol and decodes its index by matching the received multipath pattern against a channel-dependent template, but its payload unit remains the index of the single activated shift; the other cyclic shifts do not simultaneously carry independently modulated payload symbols. Because information is carried by sparse activation patterns rather than by independently modulating all available shifts, this index-modulation paradigm does not exploit the full sequence space for parallel transmission. Thus, although these designs improve support for uncoordinated access, their sparse index signaling leaves the rate--concurrency gap only partially closed.



We take a different approach: \emph{within each symbol, it maps independently modulated values onto multiple cyclic shifts of the same ZC root and superimposes the resulting sequences in the time domain}. Under an ideal channel, the zero cyclic autocorrelation between distinct shifts makes these superimposed components separable after cyclic correlation, thereby turning the cyclic-shift domain into a bank of parallel sequence sub-channels rather than a sparse index alphabet. Deploying the new waveform in underwater environments, however, remains challenging due to severe multipath and rapid channel variation. Moreover, the optimal design parameters depend on both the instantaneous channel condition and the prevailing interference level, so a static configuration cannot perform well across the diverse operating regimes encountered in practice. 

To address these challenges, we design \system, a cross-layer UWA system comprising two new components:

\para{a) \textbf{E}quidistant \textbf{ZC} \textbf{D}ivision Multiplexing (\waveform) waveform (\S\ref{sec:design}):} 
We design \waveform to activate cyclic shifts at an equidistant spacing $\tau$, leaving a guard region between adjacent sub-channels in the cyclic-correlation spectrum. Meanwhile, increasing $\tau$ reduces the number of active sub-channels but better separates their multipath responses, exposing a direct rate--robustness tradeoff.
To tolerate rapid channel variation without explicit CIR estimation, \waveform differentially encodes data across sub-channels within each symbol (i.e., intra-symbol differential modulation). Since these sub-channels share the same channel response, their relative phases remain decodable even when the response changes across symbols. At the receiver, the periodic peak structure induced by $\tau$ enables lightweight estimation of relative path weights from the cyclic-correlation spectrum, after which weighted differential decisions combine energy across paths. Then, a hard or soft demodulator converts multipath-weighted symbol probabilities into bits or bit LLRs for channel decoding.

\para{b) Concurrent MAC and Link Adaptation for EZCDM (\S\ref{sec:amc}):}
For concurrent uplinks, \system assigns each user a distinct ZC root, allowing the gateway to process overlapping transmissions with user-specific correlators without global timing coordination. 
A beacon-framed, continuously randomized access procedure admits on-demand transmissions without per-packet grants, while user-specific closed-loop power control mitigates the near--far effect. The preamble--midamble--postamble frame lets the gateway measure per-user received power, packet overlap, and intra-packet CIR similarity. It combines these observations into an effective SINR and selects a common modulation scheme that jointly configures the differential alphabet and $\tau$ for the next superframe.

\para{Implementation and Evaluation.}
We implement a custom acoustic SDR prototype and evaluate \system using channel-trace and signal-trace data in lake and diving-pool environments, together with PHY-in-the-loop network simulations of a 12-node star topology.
At the physical layer, under low-to-moderate channel variability, EZCDM achieves orders-of-magnitude lower uncoded BER than DSSS, CSS, and OFDM-DQPSK, with at least a $5$\,dB $E_b/N_0$ gain over CSS at a BER of $0.02$; under high variability the gain increases to $10$\,dB, confirming \waveform's robustness under severe channel dynamics.
In diving-pool signal-trace experiments, MS~1 and MS~2 achieve near error-free transmission in quasi-static channels and an order-of-magnitude BER reduction under highly dynamic conditions.

At the network level, \system delivers substantial throughput gains: compared to TDMA, S-ALOHA, and ZCMod (state-of-the-art concurrent system), it achieves 7.13$\times$, 2.40$\times$, and 2.02$\times$ improvements at 0.16\,packets/slot, with peak gains of 10.6$\times$, 2.71$\times$, and 2.92$\times$, respectively. The advantage decomposes into contributions from the concurrent decoding with designed on-demand access (vs.~TDMA), the concurrent decoding (vs.~S-ALOHA), and EZCDM's higher-rate robust waveform (vs.~ZCMod), together converting traditionally destructive collisions into decodable concurrent streams.

\rev{
Our main technical contributions are as follows:
\begin{itemize}
    \item We introduce EZCDM, a dense sequence-domain waveform that multiplexes independently modulated values over equidistant cyclic shifts of a ZC root, converting the shift domain from a sparse index alphabet into parallel payload sub-channels with a tunable rate--multipath-robustness tradeoff.
    \item We design an intra-symbol differential receiver that avoids explicit CIR estimation and equalization under rapidly varying channels, and a lightweight multipath-aware demodulator that estimates relative path weights directly from the cyclic-correlation spectrum to improve hard and soft decisions.
    \item We build \system, a cross-layer concurrent-access system that combines root-separated asynchronous uplinks, user-specific receive-power control, and overlap- and CIR-aware closed-loop link adaptation, and validate it through pool experiments, real-channel replay, and link- and network-level simulations.
\end{itemize}
}

\section{Background and Motivation} \label{sec:bg}

\subsection{System Model}

We consider a star network consisting of one half-duplex gateway and $K$ half-duplex registered users. Users initiate packet transmissions without per-packet scheduling or global uplink timing coordination. Let $\mathcal{U}(t)\subseteq\{1,\ldots,K\}$ denote the set of users whose packets are present at the gateway at time $t$. Their transmissions may partially or fully overlap, and the gateway seeks to identify and decode each active user.

\para{Underwater Acoustic Channel Model.}
Let \( f_c \) be the carrier frequency and \( P \) be the number of multipath components. The UWA channel is a time-varying, frequency-selective medium modeled by the time-varying baseband channel impulse response (CIR) \cite{tse2005fundamentals,bello1963characterization}
\begin{equation}
    h(t,\tau)= \sum_{p=1}^{P} G_p(t) e^{-j 2\pi f_c \tau_p(t)} \delta(\tau - \tau_p(t)),
\end{equation}
where $h(t,\tau)$ is the response at observation time $t$ to an impulse transmitted at time $t-\tau$, and $G_p(t) \in \mathbb{C}$ and $\tau_p(t)$ denote the time-varying complex path gain and instantaneous propagation delay of the $p$-th path, respectively.

Unlike narrowband RF systems where the bandwidth is negligible relative to the carrier, in UWA links the signal bandwidth can be a sizable fraction of $f_c$, so motion-induced distortion is better captured by time scaling than by a frequency shift alone. Over a short observation window, the instantaneous delay of the $p$-th path is approximated as $\tau_p(t)\approx\tau_p-a_pt$, where $\tau_p$ is the delay at the beginning of the window and $a_p=v_p/c$ is the path-dependent Doppler scale ($v_p$ is the closing velocity along the $p$-th path, defined as positive when the path length decreases). The scale $a_p$ represents the total uncompensated propagation-induced Doppler scaling.
Substituting this into the CIR yields:
\begin{equation} \label{eq:channel_wideband}
    h(t,\tau)=\sum_{p=1}^{P} G_p(t) e^{-j 2 \pi f_c (\tau_p - a_p t)} \delta(\tau - (\tau_p - a_p t)).
\end{equation}
Eq.~(\ref{eq:channel_wideband}) captures two distinct effects: the term $e^{j 2 \pi a_p f_c t}$ corresponds to conventional Doppler-induced phase rotation, while the argument of the Dirac delta, $\delta(\tau - (\tau_p - a_p t))$, dictates that physical delay taps continuously drift across the observation window, causing intra-symbol time-scaling.

\begin{figure}[tbp]
    \centering
    \subfloat[]{%
        \includegraphics[width=0.235\textwidth]{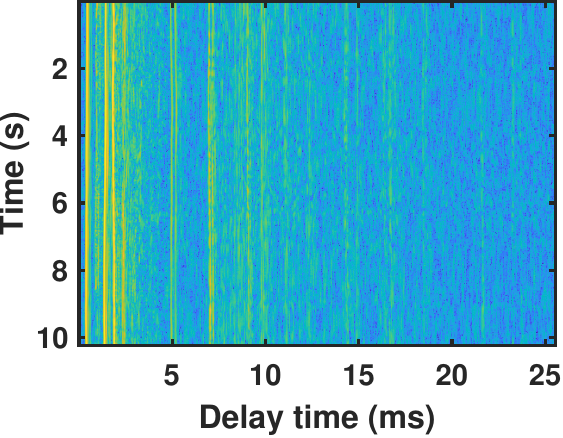}%
    }\hfill
    \subfloat[]{%
        \includegraphics[width=0.235\textwidth]{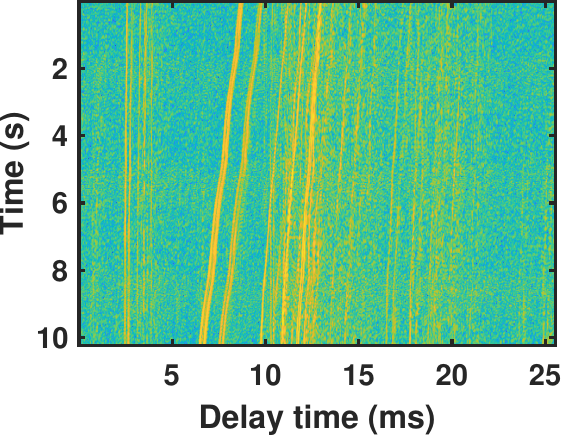}%
    }\\[1ex]
    \subfloat[]{%
        \includegraphics[width=0.235\textwidth]{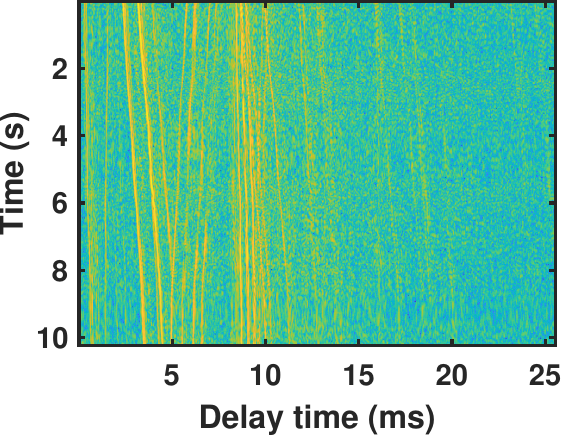}%
    }\hfill
    \subfloat[]{%
        \includegraphics[width=0.235\textwidth]{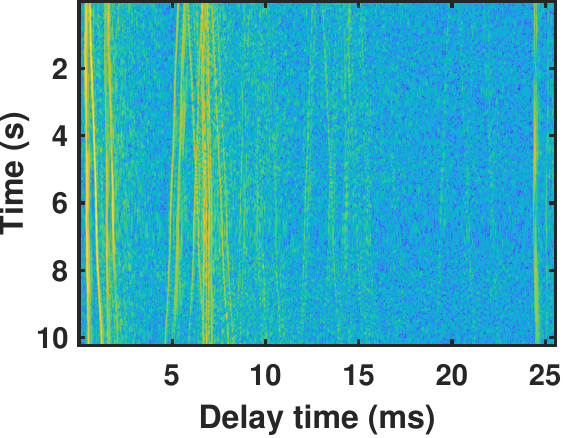}%
    }
    \caption{Underwater channel impulse responses measured in a real-world diving pool under dynamic motions: (a) hovering; (b) and (c) horizontal motions in different directions; and (d) vertical ascent and descent.}
    \label{fig_channel}
\end{figure}

Fig.~\ref{fig_channel} presents empirically measured underwater channels in $h(t,\tau)$ format ($x$-axis: delay $\tau$, $y$-axis: time $t$), collected in a diving pool under dynamic motions (experimental setup in \S\ref{sec:eval:exp}). Hovering motion (Fig.~\ref{fig_channel}(a)) exhibits pronounced multipath with minimal Doppler spread. Horizontal motions at $\approx 0.5\text{m/s}$ (Fig.~\ref{fig_channel}(b,c)) exhibit strong Doppler effects. Vertical motion (Fig.~\ref{fig_channel}(d)) shows a mixture of static and dynamic paths. These measurements highlight that multipath isolation and temporal robustness must be addressed simultaneously.

\para{Transmitter and Receiver Model.}
Let $x(t)$ be the complex-baseband packet transmit waveform. Based on the wideband CIR in Eq.~(\ref{eq:channel_wideband}), the received baseband signal in a \emph{single-user scenario} is
\begin{equation}
\resizebox{0.9\linewidth}{!}{$
\begin{aligned}
y(t) &= e^{j2\pi \nu^{\mathrm{cfo}} t}
         \int h(t,\tau)x(t-\tau)\,d\tau + w(t) \\
     &= e^{j2\pi \nu^{\mathrm{cfo}} t}
        \sum_{p=1}^{P} G_p(t)e^{-j2\pi f_c(\tau_p-a_pt)}
        \cdot x\bigl((1+a_p)t-\tau_p\bigr) + w(t).
\end{aligned}
$}
\end{equation}
Here, $\nu^{\mathrm{cfo}}$ is the carrier-frequency offset (CFO) and $w(t)$ denotes residual complex Gaussian noise after bandpass filtering. 

In an uncoordinated \emph{multi-user random access} setting, each active user $u\in\mathcal{U}(t)$ transmits $x_u(t)$, arriving with an arbitrary access delay $\Delta_u\in\mathbb{R}$. Designating user 1 as the target user ($\Delta_1=0$), the superimposed received signal is
\begin{equation}
\small
\begin{aligned}
y(t)={}&\sum_{u\in\mathcal{U}(t)} e^{j2\pi\nu_u^{\mathrm{cfo}}t}
        \sum_{p=1}^{P_u} G_{u,p}(t)
        e^{-j2\pi f_c(\tau_{u,p}-a_{u,p}t)} \\
       &\cdot x_u((1+a_{u,p})t-\Delta_u-\tau_{u,p})
        + w(t),
\end{aligned}
\end{equation}
where $P_u$, $G_{u,p}(t)$, $\tau_{u,p}$, and $a_{u,p}$ characterize the user-specific multipath channel, and each user has an independent residual CFO $\nu_u^{\mathrm{cfo}}$ common across its multipath components.

In summary, the UWA channel combines severe multipath delay spread with highly dynamic time variation, while the multi-user setting adds three coupled impairments: (1) each active user generates multiple time-scaled replicas that interleave with those of other users; (2) the path-dependent Doppler scales preclude a single per-user frequency correction from stabilizing all paths; and (3) the arbitrary access delays, combined with user-specific multipath, make the instantaneous interference pattern unpredictable across packets. Together, these challenges demand a practical waveform that preserves separability under multipath, avoids continuous coherent channel reconstruction, and tolerates asynchronous inter-user interference without sacrificing data rates.

\rev{
\subsection{Limitation of Existing Approaches}  \label{sec:motivation}

\para{Existing ZC payload waveforms remain single-shift and index-centric.}
ZC sequences are widely used for random-access timing in cellular systems~\cite{dahlman20134g,5g} and for preamble detection and Doppler estimation in UWA communications~\cite{li2017joint,tan2019preamble}. Recent designs further use ZC sequences for payload modulation~\cite{zhang2022zcnet,vangelista2023golden,zhang2025high}. In particular, ZCMod~\cite{zhang2025high} activates one cyclic shift per symbol and decodes its index by matching the received multipath correlation pattern against a channel-dependent template. Its payload unit, however, remains the index of the single activated shift; the other cyclic shifts do not simultaneously carry independently modulated payload symbols.

\para{Single-shift pattern matching does not recover multi-shift payloads.}
After ZC correlation, one activated shift produces a translated replica of the physical CIR, allowing single-shift methods to infer the shift index by locating a channel-shaped template. When multiple cyclic shifts are transmitted concurrently, the correlation output becomes a superposition of translated replicas of the same physical channel, each weighted by a payload-dependent complex coefficient. The receiver must separate these shift-domain components and decode the information carried in their complex relationships under an unknown shared channel response---a challenge that single-shift pattern matching does not address. The remaining task is therefore to structure shift-domain overlap and exploit the shared channel response so that multiple cyclic shifts can carry parallel payloads without continuous CIR reconstruction.

\para{Post-hoc Collision Resolution.}
Collision-resolution algorithms for chirp spread spectrum systems (e.g., LoRa) typically track sharp spectral, correlation, or time--frequency features to separate overlapping packets~\cite{wang2019mlora,xia2019ftrack,wang2020oct,tongcolora2020,tong2020nscale,xu2020fliplora,hu2020sclora,xu2021pyramid,xia2021pcube,shahid2021cic,chen2021aligntrack,li2022curvinglora,rathi2022tnb,wang2022decoding}. Long-delay-spread multipath broadens and overlaps these features, while path-dependent Doppler and arbitrary access delays make their locations and shapes vary across paths and users. Successive interference cancellation (SIC)~\cite{halperin2008taking} requires accurate, up-to-date channel estimates; otherwise residual errors propagate through cancellation stages. These assumptions are difficult to maintain in dynamic UWA channels, illustrating the need for a waveform that preserves separability by construction rather than relying on post-hoc collision resolution.

Together, these limitations call for a proactive waveform that preserves separability under asynchronous collisions by construction and eliminates the shared unknown channel response directly in the shift domain, as developed in Sec.~\ref{sec:design}.
}

\section{Physical-layer Design} \label{sec:design}



\subsection{Generic ZC-Shift Division Multiplexing}\label{sec:zcseq}
We first introduce a generic division multiplexing primitive based on cyclic shifts of a Zadoff--Chu (ZC) sequence.
Let $\mathbf{s}_r\in\mathbb{C}^{N}$ be a length-$N$ ZC sequence with root
$r$, where $r$ is coprime to $N$. For the odd lengths used in our design,
\begin{equation}
    s_r[n]=\exp\!\left(-j\pi r n(n+1)/N\right),
    \quad n=0,\ldots,N-1.
    \label{eq3}
\end{equation}
The sequence has constant amplitude. Let $\mathbf{\Pi}$ denote a one-sample
cyclic-shift operator, and collect all shifts of the same root as the columns of
\begin{equation}
    \mathbf{C}_r=
    [\mathbf{s}_r,\mathbf{\Pi}\mathbf{s}_r,\ldots,
    \mathbf{\Pi}^{N-1}\mathbf{s}_r].
    \label{eq4}
\end{equation}
The zero cyclic autocorrelation of a ZC sequence gives
$\mathbf{C}_r^{H}\mathbf{C}_r=N\mathbf{I}_N$. Thus, the columns of
$\mathbf{C}_r$ form orthogonal ZC-shift dimensions; they are logical
multiplexing dimensions rather than independent physical propagation
subchannels. Distinct roots have bounded cross-correlation and are assigned to
different users for random access, as described in Sec.~\ref{sec:amc}. \rev{Appendix~\ref{sec:whitevali} validates that multi-user cross-root interference is effectively whitened under asynchronous access.}

For a coefficient vector $\mathbf{x}\in\mathbb{C}^{N}$, the transmitted
time-domain block before cyclic-prefix insertion is
\begin{equation}
\begin{aligned}
    \mathbf{x}_{\mathrm{tx}}\triangleq\mathbf{u}
    &=\mathbf{C}_r\mathbf{x}=\sum_{q=0}^{N-1}x[q]\mathbf{\Pi}^{q}\mathbf{s}_r.
\end{aligned}
\label{eq:mod}
\end{equation}
Its support $\mathcal{A}\triangleq\operatorname{supp}(\mathbf{x})\subseteq\{0,\ldots,N-1\}$ determines the signaling mode. A single-active vector $\mathbf{x}=a\mathbf{e}_q$, where $\mathbf{e}_q$ is the $q$-th standard basis vector, corresponds to traditional index modulation when $q$ is selected by payload bits. A fully loaded vector satisfies $\mathcal{A}=\{0,\ldots,N-1\}$, while intermediate sparsity levels realize the equidistant design developed below.

The circulant structure of $\mathbf{C}_r$ provides an efficient FFT-domain realization $\mathbf{u}=\mathcal{F}_N^{-1}[\,\mathcal{F}_N(\mathbf{s}_r)\odot\mathcal{F}_N(\mathbf{x})\,]$, where $\mathcal{F}_N$ denotes the $N$-point DFT. Since $\mathcal{F}_N(\mathbf{s}_r)$ is fixed and can be precomputed, the cost is reduced to $\mathcal{O}(N\log N)$.

\rev{
Fig.~\ref{fig:modulator} illustrates the EZCDM modulator structure. Payload bits are mapped to a coefficient vector $\mathbf{x}$ with structured sparse support (the equidistant design is developed in \S\ref{subsec:ezcdm}), where each active coefficient is differential-PSK modulated. The coefficient vector is then multiplied by the precomputed ZC shift matrix $\mathbf{C}_r$, producing the time-domain block $\mathbf{u}$. Equivalently, this step can be implemented via the FFT-domain realization in the preceding paragraph.

\begin{figure}[t]
    \centering
    \includegraphics[width=0.45\textwidth]{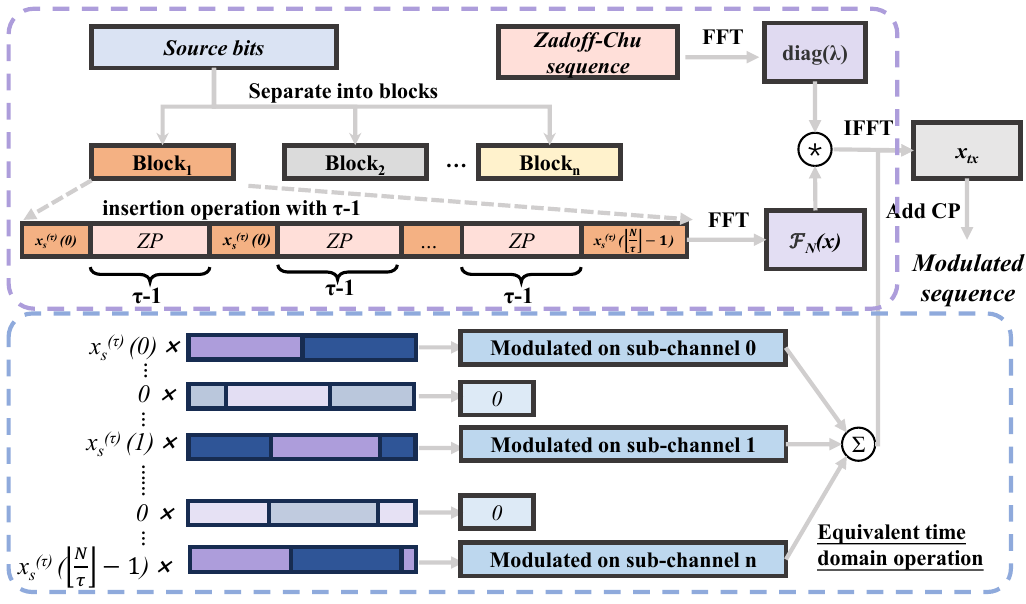}
    \caption{EZCDM modulator structure: payload bits are mapped to coefficients $\mathbf{x}$ with equidistant sparse support, then transformed to the time-domain block $\mathbf{u}$ via the ZC shift matrix $\mathbf{C}_r$.}
    \label{fig:modulator}
\end{figure}
}

Given a received block $\mathbf{y}\in\mathbb{C}^{N}$, the receiver projects it
onto all $N$ cyclic shifts. We define the resulting vector as the
\emph{ZC-shift-domain spectrum}:
\begin{equation}
\begin{aligned}
    \mathbf{z}
    &\triangleq\frac{1}{N}\mathbf{C}_r^{H}\mathbf{y}=\frac{1}{N}\mathcal{F}_N^{-1}\!\left[
    \mathcal{F}_N(\mathbf{s}_r)^{*}\odot
    \mathcal{F}_N(\mathbf{y})\right],
\end{aligned}
\label{eq:demod}
\end{equation}
which can be computed in $\mathcal{O}(N\log N)$ via one FFT, one element-wise multiplication, and one IFFT. Its $q$-th bin is the matched cyclic-correlation coefficient for the $q$-sample shift of $\mathbf{s}_r$ and serves as the common front-end for all subsequent EZCDM demodulation.


\begin{figure*}[t!]
    \centering
    \subfloat[Single-active ($\tau=N=257$)]{%
        \includegraphics[width=0.232\textwidth]{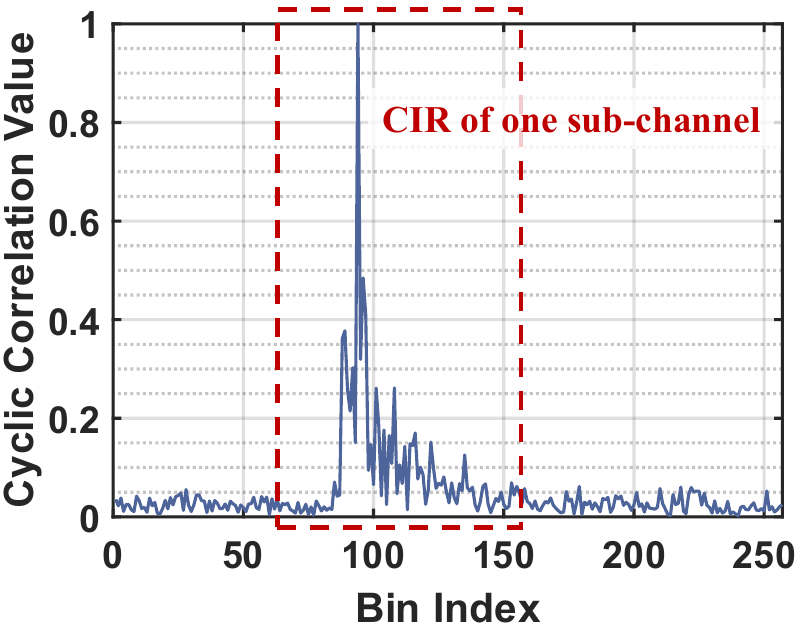}%
    }\hfill
    \subfloat[Full-load ($\tau=1$)]{%
        \includegraphics[width=0.232\textwidth]{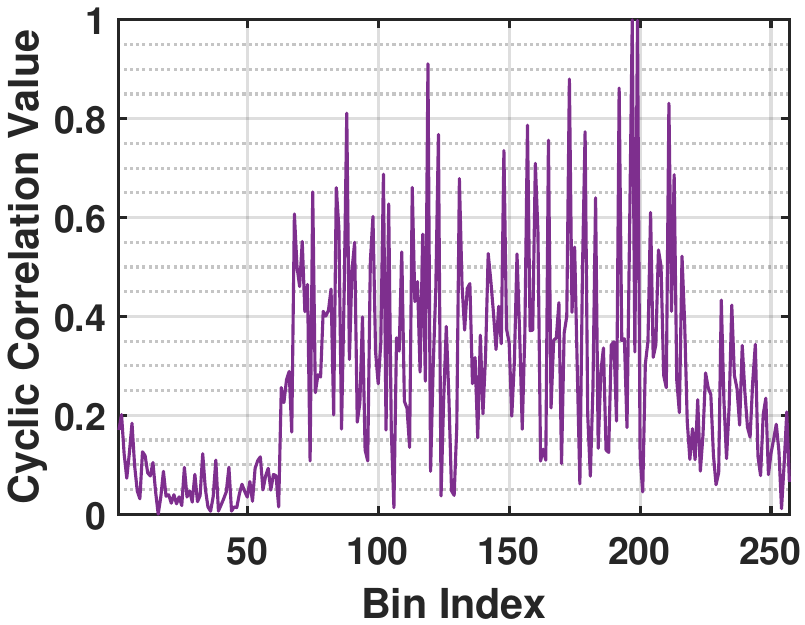}%
    }\hfill
    \subfloat[Equidistant ($\tau=17$)]{%
        \includegraphics[width=0.232\textwidth]{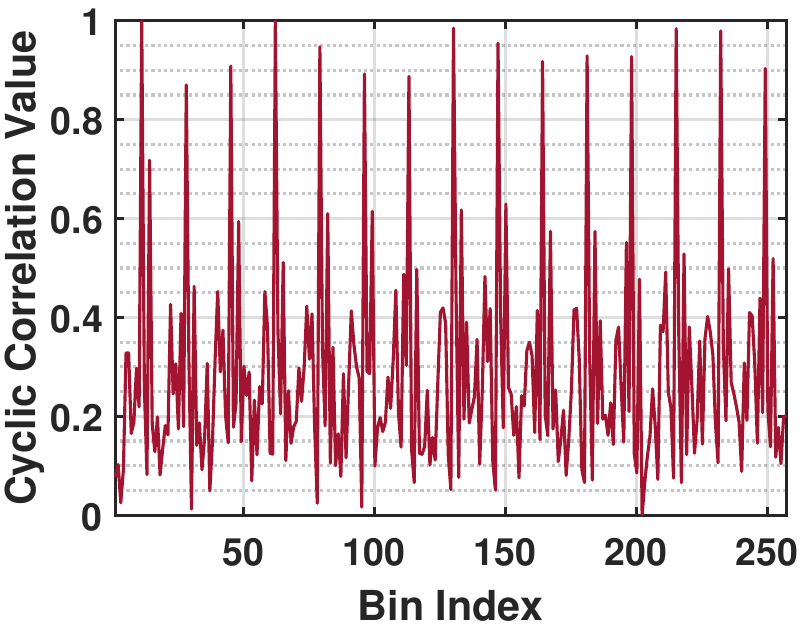}%
    }\hfill
    \subfloat[Folded path profile for (c)]{%
        \includegraphics[width=0.232\textwidth]{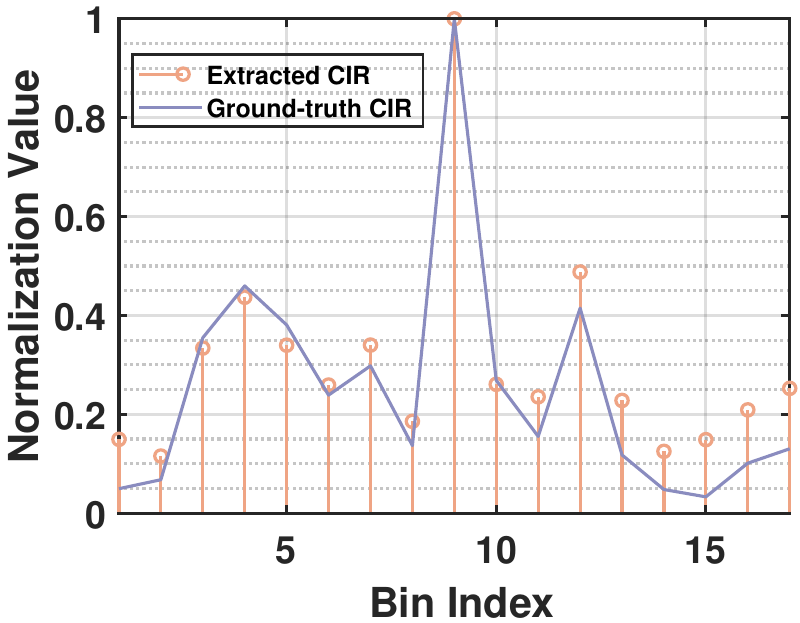}%
    }
    \caption{Normalized ZC-shift-domain spectra for $N=257$ after replay
    through a representative measured UWA channel. A single active coefficient
    produces one CIR replica; full loading causes severe overlap; equidistant
    activation separates the dominant replicas; folding those replicas
    estimates the channel profile.}
    \label{fig:cyc_pass_example}
    \vspace{-0.15in}
\end{figure*}

\subsection{Equidistant ZCDM over Multipath UWA Channels}
\label{sec:eocdm}\label{subsec:ezcdm}\label{sec:ezcdm_demod}

\subsubsection{CP-Aided Transmission}

UWA channels exhibit long multipath delay spreads, causing inter-symbol interference between consecutive waveform blocks. We therefore first prepend a cyclic prefix (CP) to each block.
Let $\mathbf{h}=[h[0],\ldots,h[L_h-1]]^{T}\in\mathbb{C}^{L_h}$ denote the
sampled CIR of length $L_h\leq N$ with maximum excess delay $D_h=L_h-1$. For each length-$N$ block $\mathbf{u}$, the transmitter prepends a CP of length $N_{\mathrm{cp}}\geq D_h$. After timing alignment and CP removal, the retained $N$ samples follow a circular-convolution model. \rev{Appendix~\ref{appendix:cp_window} analyzes the robustness of this model to residual receive-window displacement.}

Define the zero-padded channel vector $\widetilde{\mathbf{h}}=[\mathbf{h}^{T},\mathbf{0}_{N-L_h}^{T}]^{T}$ and let $\mathbf{H}_{c}$ be the circulant matrix with $\widetilde{\mathbf{h}}$ as its first column. The received block is $\mathbf{y}=\mathbf{H}_{c}\mathbf{C}_{r}\mathbf{x}+\mathbf{v}$. Since circulant matrices commute, applying Eq.~\eqref{eq:demod} gives the received spectrum
\begin{equation}
    \mathbf{z}=\mathbf{H}_{c}\mathbf{x}+\mathbf{v}_{z}
      =\widetilde{\mathbf{h}}\circledast\mathbf{x}+\mathbf{v}_{z},
    \label{eq:ezcdm_channel}
\end{equation}
where $\mathbf{v}_{z}\triangleq\frac{1}{N}\mathbf{C}_{r}^{H}\mathbf{v}$ and $\circledast$ denotes length-$N$ circular convolution. Thus, every active coefficient in $\mathbf{x}$ produces a shifted replica of the CIR in the ZC-shift-domain spectrum. Eq.~\eqref{eq:ezcdm_channel} establishes the shift-domain observation model used by the remainder of the receiver.

\subsubsection{Equidistant Activation}

EZCDM chooses the sparse support at equidistant indices. For spacing $\tau\in\{1,\ldots,N\}$, we use
\begin{equation}
    K=\left\lfloor\frac{N}{\tau}\right\rfloor,
    \qquad q_{\ell}=\ell\tau,
    \quad \ell=0,\ldots,K-1,
    \label{eq:ezcdm_active_indices}
\end{equation}
and construct the coefficient vector as
\begin{equation}
    x[q]=
    \begin{cases}
        A_{\tau}a[\ell], & q=q_{\ell},\quad \ell=0,\ldots,K-1,\\
        0, & \text{otherwise},
    \end{cases}
    \label{eq:ezcdm_support}
\end{equation}
where $a[\ell]\in\mathbb{C}$ is a unit-modulus coefficient and $A_{\tau}$ is a common amplitude-normalization factor. Since $\mathbf{x}$ has $K$ nonzero entries of magnitude $A_\tau$, the useful block energy is $E_u=\|\mathbf{u}\|^2=N\|\mathbf{x}\|^2=NK A_\tau^2$, giving $A_\tau=\sqrt{E_u/(NK)}$.

Let $D_{\epsilon}$ be the span containing the significant taps of the CIR under an energy threshold $\epsilon$. A sufficient no-overlap condition for the replicas is
\begin{equation}
    D_{\epsilon}\leq
    \min\!\left\{\tau,\,N-(K-1)\tau\right\}.
    \label{eq:ezcdm_spacing_condition}
\end{equation}
The second term accounts for the circular gap between the last and first active indices. In practice this is a design target: increasing $\tau$ suppresses replica overlap and increases the energy per active dimension, but decreases $K$ and hence the carried bits per block.

Fig.~\ref{fig:cyc_pass_example} illustrates this progression on a measured channel trace with $N=257$. At $\tau=N$ (i.e., \cite{zhang2025high}), the shift-domain spectrum exposes one shifted CIR replica. At $\tau=1$, the replicas are densely superimposed and are irresolvable. The intermediate choice $\tau=17$ separates the dominant paths while retaining multiple active ZC-shift dimensions.

\subsubsection{Intra-Symbol Differential Modulation}

For a significant path offset $d$ and nonoverlapping replicas, Eq.~\eqref{eq:ezcdm_channel} reduces to
\begin{equation}
    z[(q_{\ell}+d)\bmod N]
    \approx A_{\tau}h[d]a[\ell]
    +v_z[(q_{\ell}+d)\bmod N].
    \label{eq:ezcdm_per_path_obs}
\end{equation}
Hence, for a fixed $d$, all active dimensions within one waveform block share approximately the same complex path gain. This motivates intra-symbol differential encoding, which avoids explicit estimation of the phase of $h[d]$.

Let $M$ be a power-of-two PSK order and $B=\log_2 M$. For the $B$ bits $\{b_{\ell,b}\}_{b=0}^{B-1}$ assigned to active dimension $\ell$, define
\begin{equation}
    u_{\ell}=\sum_{b=0}^{B-1}2^{b}b_{\ell,b},
    \qquad
    c[\ell]=\exp\!\left(j\frac{2\pi u_{\ell}}{M}\right),
    \label{eq:modmpsk}
\end{equation}
for $\ell=1,\ldots,K-1$. The transmitted coefficients are
\begin{equation}
    a[0]=1,\qquad
    a[\ell]=a[\ell-1]c[\ell],
    \quad \ell=1,\ldots,K-1.
    \label{eq:intra_diff_encoding}
\end{equation}
The first active dimension is a differential reference; one waveform block carries $(K-1)B$ payload bits. From two adjacent observations on the same path, the receiver forms
\begin{align}
    v_{\ell,d}
    &\triangleq z[(q_{\ell}+d)\bmod N]
    z^{*}[(q_{\ell-1}+d)\bmod N] \notag\\
    &\approx A_{\tau}^{2}|h[d]|^{2}c[\ell]+\eta_{\ell,d},
    \label{eq:intra_diff_obs}
\end{align}
where $\eta_{\ell,d}$ collects noise and residual replica interference. The unknown path phase is eliminated without assuming that different physical paths share the same phase.

\begin{figure}[t!]
    \centering
    \subfloat[$\tau=17$, 1st block]{%
        \includegraphics[width=0.156\textwidth]{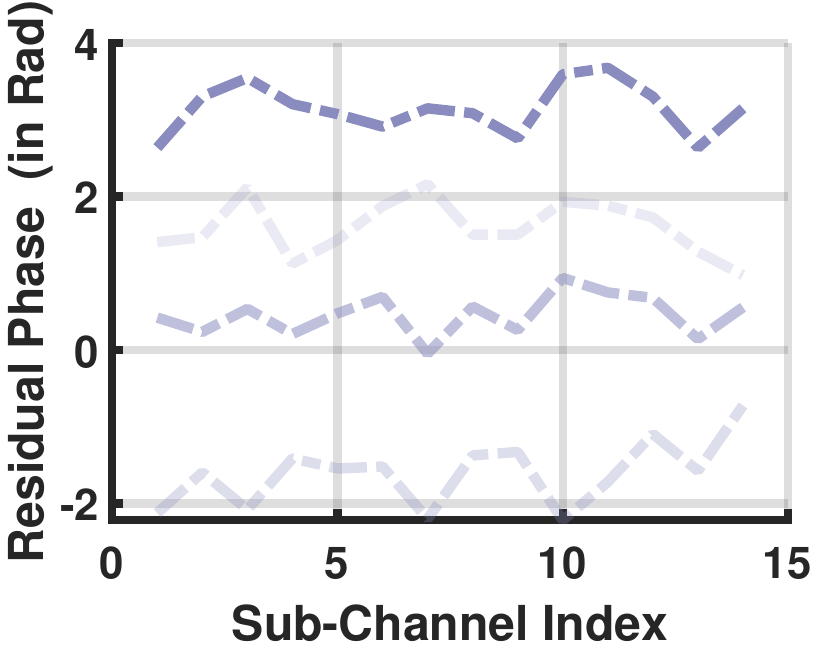}%
    }\hfill
    \subfloat[$\tau=17$, 2nd block]{%
        \includegraphics[width=0.156\textwidth]{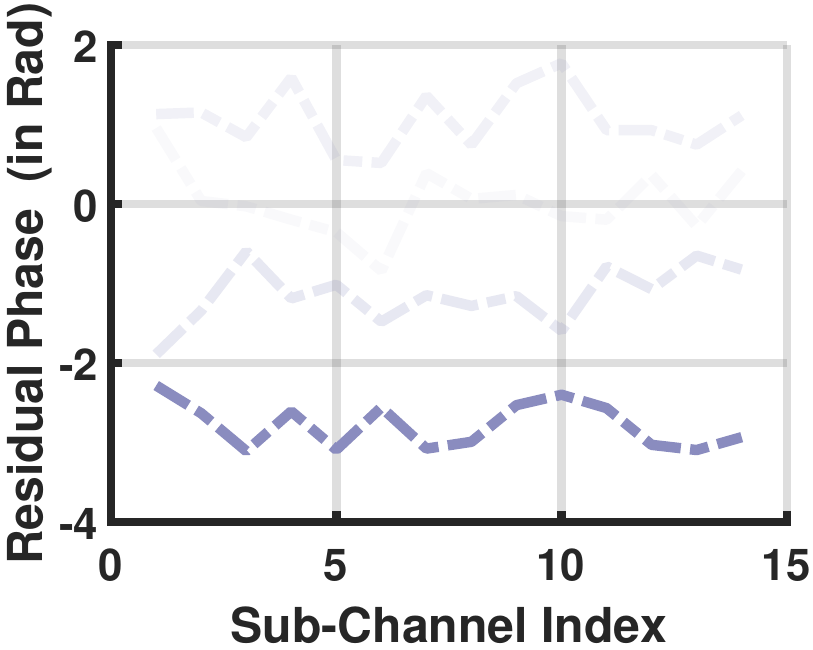}%
    }\hfill
    \subfloat[$\tau=17$, 15th block]{%
        \includegraphics[width=0.156\textwidth]{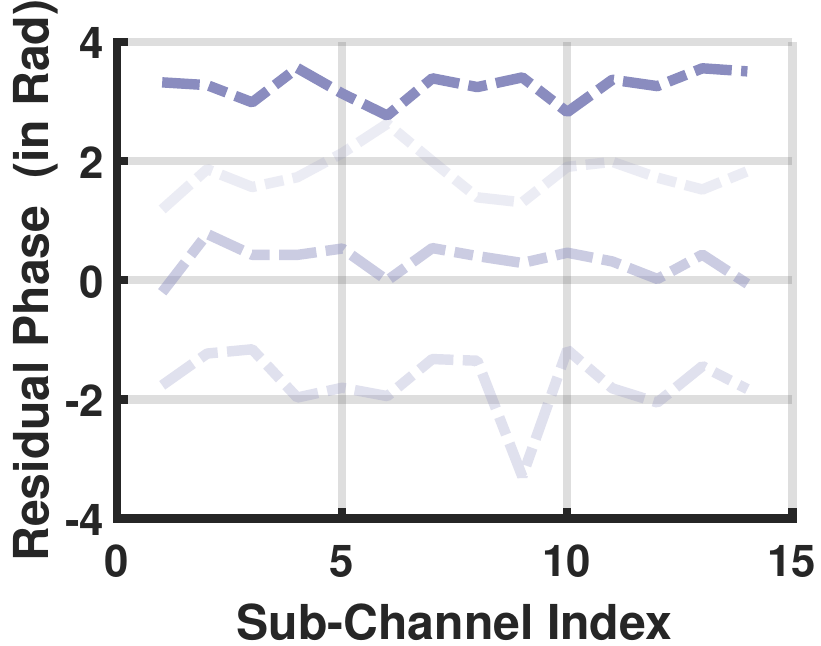}%
    }\\[1ex]
    \subfloat[$\tau=4$, 1st block]{%
        \includegraphics[width=0.156\textwidth]{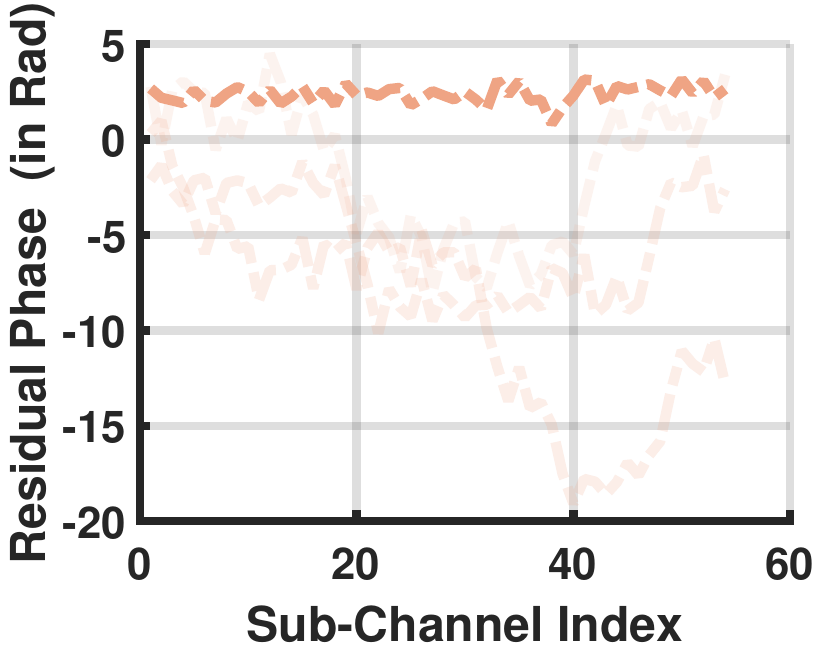}%
    }\hfill
    \subfloat[$\tau=4$, 2nd block]{%
        \includegraphics[width=0.156\textwidth]{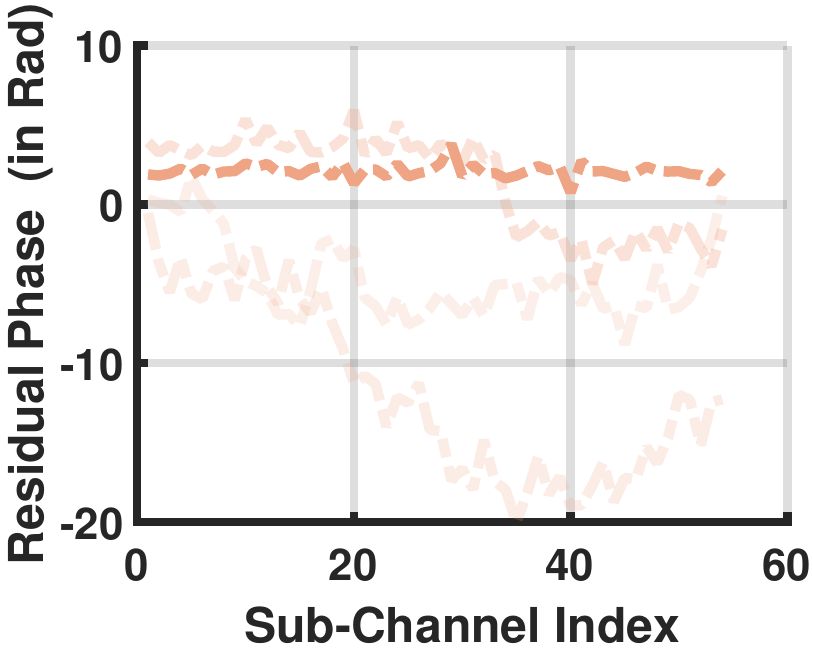}%
    }\hfill
    \subfloat[$\tau=4$, 15th block]{%
        \includegraphics[width=0.156\textwidth]{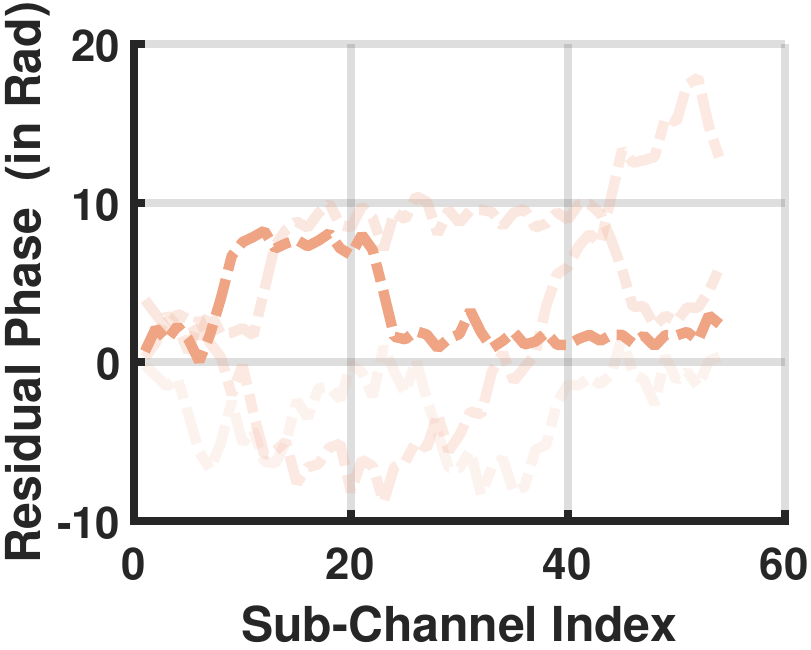}%
    }
    \caption{ZC-shift-domain phases at representative strong path offsets for
    waveform blocks with $N=257$ replayed through measured UWA channels. Darker curves denote stronger paths. With $\tau=17$, each fixed path offset has a nearly common phase across the active dimensions within a block; the smaller spacing $\tau=4$ exhibits stronger replica interference.}
    \label{fig:measured_path_phase}
\end{figure}

Fig.~\ref{fig:measured_path_phase} empirically checks the common-path approximation in Eq.~\eqref{eq:ezcdm_per_path_obs}. For $\tau=17$, the phase at each selected path offset is stable across active dimensions within successive blocks, although it changes between blocks. At $\tau=4$, stronger replica overlap produces visibly less regular observations.

\subsubsection{Multipath-Aware Demodulation}\label{sec:decoding}

Using only the strongest path discards energy carried by other replicas. EZCDM instead folds the ZC-shift-domain energy modulo $\tau$ to estimate a path profile:
\begin{equation}
    \beta[d]=\sum_{\ell=0}^{K-1}
    \left|z[(q_{\ell}+d)\bmod N]\right|^{2},
    \quad d=0,\ldots,\tau-1.
    \label{eq:sliding_energy}
\end{equation}
This accumulation over $\ell$ estimates blockwise path reliability. Significant path offsets are retained as
\begin{equation}
    \widehat{\mathcal{D}}
    =\left\{d:\beta[d]\geq\rho\max_{d'}\beta[d']\right\},
    \qquad 0<\rho\leq1.
    \label{eq:significant_paths}
\end{equation}

For each differential-symbol index $\ell=1,\ldots,K-1$ and path offset $d$, we normalize the differential statistic to reduce path-amplitude bias,
\begin{equation}
    \bar v_{\ell,d}=
    \frac{v_{\ell,d}}
    {|z[(q_{\ell}+d)\bmod N]|
     |z[(q_{\ell-1}+d)\bmod N]|}. 
    \label{eq:normalized_diff_stat}
\end{equation}
For each $\ell$ and differential modulation candidate $c_m=\exp(j2\pi m/M)$, the multipath-combined metric is
\begin{equation}
    \Lambda_m[\ell]=-
    \sum_{d\in\widehat{\mathcal{D}}}w_d
    \left|\bar v_{\ell,d}-c_m\right|^{2},
    \quad
    w_d=\frac{\beta[d]}{\sum_{d'\in\widehat{\mathcal{D}}}\beta[d']}.
    \label{ori_p}
\end{equation}
The hard output is $\widehat m_{\ell}=\arg\max_{m\in\{0,\ldots,M-1\}}\Lambda_m[\ell]$, and the same metrics can be converted to bit LLRs for soft decoding \rev{(see Appendix~\ref{appendix:llr})}.

\section{Link-layer Design}\label{sec:amc}

\subsection{Design Overview}


The multiuser reception capability of \waveform creates three link-layer challenges, which we address while preserving asynchronous random access:

\para{$\rhd$ Near--far effect.}
Different ZC roots make concurrent packets separable but do not eliminate the power disparity caused by heterogeneous path losses. We therefore use \emph{user-specific closed-loop power control}: the gateway measures each user's uplink power from its assigned-root known symbols and returns a coarse adjustment command.

\para{$\rhd$ Half-duplex constraint.}
Returning power commands requires a predictable downlink opportunity, yet a half-duplex gateway cannot broadcast control while receiving. We therefore use \emph{beacon-framed random access}: each beacon is followed by an uplink interval in which users independently apply continuous random backoffs, protecting control without per-packet grants, slots, or precise synchronization.

\para{$\rhd$ Dynamic operating regime.}
Channel evolution and traffic variation change the required robustness and interference level. We therefore use \emph{overlap- and channel-aware link adaptation} to select a common modulation scheme (MS) from recent uplink observations. 
\rev{The overall design is conservative: a worst-case SINR estimate is chosen to satisfy the target PER on every observed link.}

\begin{figure*}
    \centering
    \includegraphics[width=0.9\textwidth]{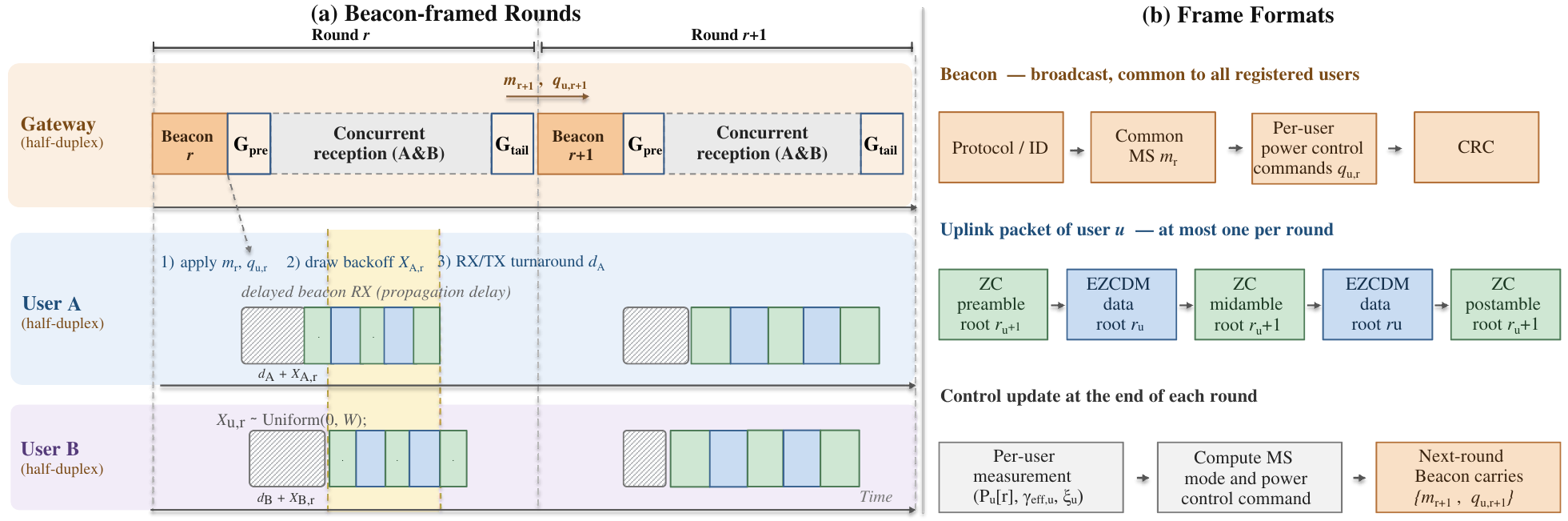}
    \caption{Example of beacon-framed random access and adopted frame formats in \system.}
    \label{fig_framework}
\end{figure*}

We present a beacon-framed, half-duplex random-access protocol exampled in Fig.~\ref{fig_framework}, which uses two frame types. The beacon, transmitted via the ZCMod frame with a dedicated root, carries the common MS $m_r$ and one power-control command per registered user $p_{u,r}$, with no per-packet grant, ACK, or retry state. The uplink data packet contains a preamble, two EZCDM data regions separated by a midamble, and a postamble; for user $u$, the preamble, midamble, and postamble are known ZC symbols (root $r_u+1$), while data regions use \waveform (root $r_u$). These known symbols jointly support user identification, packet-boundary detection, receive-power measurement, and CIR-similarity estimation.

In one round, the gateway broadcasts a beacon carrying the common MS and per-user power commands, then switches to receive mode. Eligible users apply their command and the advertised mode, take a queue snapshot, draw a continuous random backoff, and transmit at most one packet. The gateway runs per-user ZC correlators and \waveform decoders in parallel, recording arrival intervals and extracting per-user power, CIR-similarity, and effective SINR estimates. During the reserved tail interval, it updates the power-control EWMAs, computes the per-user commands for the next beacon, and applies the rules to select the common MS for the following superframe. 
The following subsections detail these designs.

%

\subsection{Beacon-Framed Random Access}

The gateway beacon is transmitted using ZCMod---a single-shift ZC index modulation---with a dedicated root known to all registered users. It provides a coarse timing anchor and a protected opportunity to broadcast common control. Eligible users independently choose continuous backoffs within a fixed window; their packets may still overlap asynchronously. The superframe bounds arrivals before the next beacon without slots, grants, timing advance, or precise synchronization.

\subsubsection{Superframe Timing}

Let $t_r$ be the gateway beacon start time and $T_b$ its airtime. The single-trip propagation delay to user $u$ is $\tau_u$. After receiving the beacon, user $u$ requires a local turnaround $d_u$ for processing, RX--TX switching, and a fixed holdoff. Its maximal timing error is $T_{\rm clk}$.

We define an RTT-inclusive pre-budget
\begin{equation}
    G_{\rm pre}\geq
    \max_{u\in\mathcal U}\left(2\tau_u+d_u\right)
    +T_{\rm clk}, 
    \label{eq:llc_pre_budget}
\end{equation}
and a tail budget for computation and RX--TX switching,
\begin{equation}
    G_{\rm tail}=T_{\rm ctrl}^{\rm GW}
    +T_{\rm sw}^{\rm GW,RX\rightarrow TX}.
    \label{eq:llc_tail_budget}
\end{equation}
Let $W$ be the continuous-backoff window, $\mathcal M$ the set of supported MSs, and $T_m$ the airtime of one uplink packet under mode $m$. Reserving $T_{\max}=\max_{m\in\mathcal M}T_m$ yields the fixed superframe duration
\begin{equation}
    T_{\rm SF}=T_b+G_{\rm pre}+W+T_{\max}+G_{\rm tail},
    \qquad t_{r+1}=t_r+T_{\rm SF}.
    \label{eq:llc_superframe_duration}
\end{equation}

Suppose an eligible user draws $X_{u,r}\in[0,W]$. Its packet-arrival start at the gateway is
\begin{equation}
    a_{u,r}=t_r+T_b+2\tau_u+d_u+X_{u,r}, 
    \label{eq:llc_gateway_arrival_time}
\end{equation}
with end $b_{u,r}=a_{u,r}+T_{m_r}$. From Eqs.~\eqref{eq:llc_pre_budget}--\eqref{eq:llc_superframe_duration}, $b_{u,r}\leq t_{r+1}-G_{\rm tail}<t_{r+1}$. Hence every legal packet reaches the gateway before the next beacon; propagation-delay differences naturally contribute to the arrival skew without timing advance.

\subsubsection{Beacon-Triggered Randomized Transmission}

Let $t^{\mathrm{B}}_{u,r}$ be the local time at which user $u$ completely receives and verifies beacon $r$. Eligible contenders are
\begin{equation}
    \mathcal{A}_r = \left\{u\in\mathcal{U}:Q_u
    \bigl(t^{\mathrm{B}}_{u,r}\bigr)>0\right\},
    \label{eq:contender_set}
\end{equation}
where $Q_u(t)$ is the queue length at time $t$, so the instantaneous transmitter count varies across superframes.

After verifying beacon $r$, user $u$ applies its power-control command, and installs the advertised common mode $m_r$.
Then, if the queue is empty, it remains silent. Otherwise, it locks one head-of-line packet and draws $X_{u,r}\sim\operatorname{Uniform}(0,W)$. After the mandatory turnaround, it waits $X_{u,r}$ and transmits once. Packets arriving after the snapshot wait for a subsequent beacon, limiting each user to one attempt per superframe.

Because backoffs are continuously distributed, different users may overlap partially or almost completely; the gateway separates user-specific observations using the assigned-root receiver in Sec.~\ref{sec:design}. A user that fails to verify the beacon remains silent, avoiding stale timing or an obsolete MS. After transmission, the packet leaves the link-layer queue regardless of the decode outcome---no implicit ARQ is created.

\subsection{Per-User Power Control}
\label{sec:power_control}

The gateway closes one low-rate power-control loop per user, estimating received power from assigned-root known symbols, smoothing observations, and comparing against a common target. The next beacon returns one quantized \texttt{UP}, \texttt{DOWN}, or \texttt{HOLD} command per user.

\subsubsection{User-Specific Receive-Power Estimation}

The preamble, midamble, and postamble of user $u$ use root $r_u+1$ and are separable from the data root $r_u$. For known symbol $s\in\mathcal{S}_{u,r}$ detected in round $r$, let $\mathbf{z}_{u,s}$ be its ZC-shift-domain spectrum and $\widehat{\mathcal D}_{u,s}$ the significant CIR offsets from Eq.~\eqref{eq:significant_paths}. We estimate the user-specific received signal power as
\begin{equation}
    \widehat P_{u,s}
    =\Bigl[
      \sum_{d\in\widehat{\mathcal D}_{u,s}}
      \left|z_{u,s}[d]\right|^{2}
      -|\widehat{\mathcal D}_{u,s}|\widehat\sigma_{u,s}^{2}
     \Bigr]_{+},
    \label{eq:pc_symbol_power}
\end{equation}
where $\widehat\sigma_{u,s}^{2}$ is estimated from off-profile bins. Summing significant path energies makes the estimate insensitive to which path dominates, while noise-floor subtraction prevents weak users from appearing artificially strong.

The gateway aggregates multiple valid known symbols in the logarithmic domain:
\begin{equation}
    \widehat p_{u,r}
    =\frac{1}{|\mathcal{S}_{u,r}^{\rm val}|}
      \sum_{s\in\mathcal{S}_{u,r}^{\rm val}}
      10\log_{10}\!\bigl(\widehat P_{u,s}/P_{\rm ref}\bigr),
    \label{eq:pc_packet_power}
\end{equation}
where $\mathcal{S}_{u,r}^{\rm val}$ contains valid known-symbol observations. The gateway maintains the exponentially weighted state
\begin{equation}
    \bar p_{u,r}=(1-\alpha_p)\bar p_{u,r^-}
                  +\alpha_p\widehat p_{u,r},
    \qquad 0<\alpha_p\leq1,
    \label{eq:pc_ewma}
\end{equation}
whenever a valid observation is available, where $r^-$ is the most recent preceding round in which user $u$ was observed. Silent rounds preserve the estimate unchanged; before the first valid observation, the state is unavailable.

\subsubsection{Quantized Closed-Loop Control}

Let $p^\star$ be the desired received power in dB and $\delta_p>0$ a deadband. For a user validly observed in round $r$, the gateway generates
\begin{equation}
q_{u,r+1}=
\begin{cases}
\texttt{UP},   & \bar p_{u,r}<p^\star-\delta_p,\\
\texttt{DOWN}, & \bar p_{u,r}>p^\star+\delta_p,\\
\texttt{HOLD}, & \text{otherwise}.
\end{cases}
\label{eq:pc_command}
\end{equation}
If no valid observation exists, $q_{u,r+1}=\texttt{HOLD}$. The command vector is inserted into beacon $r+1$, making the one-superframe feedback delay explicit.

Let $g_{u,r}$ denote user $u$'s transmit gain in dB and $\Delta_p$ the control step. After receiving beacon $r+1$, the user applies
\begin{equation}
g_{u,r+1}=
\begin{cases}
\left[g_{u,r}+\Delta_p\right]_{g_{\min}}^{g_{\max}},
    & q_{u,r+1}=\texttt{UP},\\
\left[g_{u,r}-\Delta_p\right]_{g_{\min}}^{g_{\max}},
    & q_{u,r+1}=\texttt{DOWN},\\
g_{u,r}, & q_{u,r+1}=\texttt{HOLD}.
\end{cases}
\label{eq:pc_gain_update}
\end{equation}
The updated gain applies to the entire next uplink packet. A user missing the beacon retains its current gain. 

\subsection{Overlap- and Channel-Aware Link Adaptation}
\label{sec:link_adaptation}

The gateway summarizes each observed link by its worst-case effective SINR and within-packet CIR similarity, queries an offline PER LUT, and feeds back one common MS with conservative fallback and asymmetric hysteresis.

\subsubsection{Modulation Schemes}

We expose four modulation schemes: MS~1=$(\mathrm{DBPSK},\tau=17)$, MS~2=$(\mathrm{DBPSK},\tau=9)$, MS~3=$(\mathrm{DQPSK},\tau=17)$, and MS~4=$(\mathrm{DQPSK},\tau=9)$. Larger spacing reduces ZC-shift-domain replica overlap and allocates more energy per active coefficient but lowers the number of subchannels. DBPSK trades one bit per differential symbol for greater phase-variation robustness. The best choice depends jointly on interference and channel variation.

\subsubsection{CIR Similarity}

The preamble, midamble, and postamble provide three user-specific CIR snapshots per detected packet. For two snapshots separated by $\Delta t$, we use the magnitude of their normalized complex inner product,
\begin{equation}
    \xi_u(t,t+\Delta t)=
    \left|
    \frac{\sum_{i=0}^{L-1}\widehat h_u(t,i)
    \widehat h_u^*(t+\Delta t,i)}
    {\sqrt{\sum_{i=0}^{L-1}|\widehat h_u(t,i)|^2}
     \sqrt{\sum_{i=0}^{L-1}|\widehat h_u(t+\Delta t,i)|^2}}
    \right|,
    \label{eq:cir_pair_similarity}
\end{equation}
where $\widehat h_u(t,i)$ is tap $i$ of user $u$'s estimated CIR over length $L$. Let $t_{u,r}^{\rm pre}$, $t_{u,r}^{\rm mid}$, $t_{u,r}^{\rm post}$ denote the three reference-symbol times. We summarize within-packet variation as
\begin{equation}
    \widehat\xi_u[r]=\frac{1}{2}
    \bigl(\xi_u^{\rm pre,mid}[r]+\xi_u^{\rm mid,post}[r]\bigr),
    \label{eq:cir_mean_similarity}
\end{equation}
evaluated only when all three reference symbols pass root-assignment and CIR-quality checks. A smaller value indicates faster channel evolution or stronger multipath-profile variation; the metric does not require a successful payload CRC.

\subsubsection{Worst-Case SINR Estimation}

Let $\mathcal O_r$ be the set of users with valid interval and power estimates in round $r$. For each detected user $u$, the gateway obtains its arrival interval $[a_{u,r},b_{u,r}]$ from reference symbols. All other users in $\mathcal O_r$ whose intervals overlap with $[a_{u,r},b_{u,r}]$ are treated as interferers:
\begin{equation}
    \mathcal I_{u,\max}[r] = \bigl\{v\neq u \mid
    [a_{v,r},b_{v,r}]\cap [a_{u,r},b_{u,r}]\neq\varnothing\bigr\}.
    \label{eq:max_overlap_set}
\end{equation}
Since multi-user interference dominates noise in the concurrent-access regime, the effective per-user SINR is approximated by the power ratio against the set of all overlapping interferers:
\begin{equation}
    \widehat\gamma_{{\rm eff},u}[r] \approx
    \frac{\widehat P_u[r]}
    {\sum_{v\in\mathcal I_{u,\max}[r]}\widehat P_v[r]},
    \label{eq:worst_case_sinr}
\end{equation}
where $\widehat P_u[r]$ is from Sec.~\ref{sec:power_control}. This formulation is deliberately conservative: it replaces the time-varying interference pattern with the worst case where all overlapping interferers are active simultaneously. The subsequent common-MS selection further protects the user with the smallest $\widehat\gamma_{{\rm eff},u}[r]$, yielding a doubly conservative design.

\subsubsection{Common-MS Selection and Feedback}

Offline, we superimpose \waveform packets over an SINR grid for channels spanning the observed CIR-similarity range and build the MS--SINR--CIR-Similarity--PER look-up table $\widehat{\mathrm{PER}}(m,\gamma,\xi)$ \rev{(see Appendix~\ref{appendix:lut} for construction details)}. For a candidate mode and observation pair, predicted goodput is
\begin{equation}
    G(m;\gamma,\xi)=R_m
    \bigl[1-\widehat{\mathrm{PER}}(m,\gamma,\xi)\bigr],
    \label{eq:la_goodput}
\end{equation}
where $R_m$ is the coded PHY rate.

All packets in superframe $r$ use the common mode $m_r$ announced in beacon $r$. After reception, define the valid adaptation set
\begin{equation}
    \mathcal V_r=\bigl\{u:\widehat\gamma_{{\rm eff},u}[r]
    \text{ and }\widehat\xi_u[r]\text{ are both valid}\bigr\}.
    \label{eq:valid_la_set}
\end{equation}
When $\mathcal V_r\neq\varnothing$, the gateway first forms the feasible modes satisfying the target PER on every valid link,
\begin{equation}
    \mathcal F_r=\Bigl\{m\in\mathcal M:
    \max_{u\in\mathcal V_r}
    \widehat{\mathrm{PER}}\!\bigl(m,
    \widehat\gamma_{{\rm eff},u}[r],\widehat\xi_u[r]\bigr)
    \leq\varepsilon_{\rm PER}\Bigr\},
    \label{eq:la_feasible_set}
\end{equation}
with $\varepsilon_{\rm PER}=0.1$. If $\mathcal F_r\neq\varnothing$, the candidate maximizes the worst-link goodput:
\begin{equation}
    m_r^{\star}\in\arg\max_{m\in\mathcal F_r}
    \min_{u\in\mathcal V_r}
    G\!\bigl(m;\widehat\gamma_{{\rm eff},u}[r],
    \widehat\xi_u[r]\bigr),
    \label{eq:common_ms_selection}
\end{equation}
with ties broken by smaller worst-link PER, then by smaller PHY rate. If $\mathcal F_r=\varnothing$, the gateway selects the mode minimizing the maximum predicted PER.




\section{Performance Evaluation} \label{sec:eval}

Our evaluation aims to answer the following questions: 
(1) Can \waveform sustain reliable high-rate communication over dynamic UWA channels without explicit CIR estimation or equalization?
(2) Can the closed-loop power-control mechanism reduce near--far imbalance in a two-user pool experiment?
and (3) Can the complete \system system convert \waveform's PHY rate and concurrent-decoding capability into higher network throughput than both non-concurrent protocols and representative concurrent-access schemes?

\subsection{Evaluation Setup}\label{sec:eval:exp}

\subsubsection{Methodology.} 
We adopt a layered evaluation methodology.
At the physical layer, we employ both channel-trace-driven and signal-trace-driven approaches: the former replays measured CIRs via the replay-filter method~\cite{Otnes2013replay}, while the latter replays recorded passband signals directly.
Both operate in a single-user setting, which is representative because multi-user cross-root interference is effectively whitened.

At the network layer, we conduct system-level simulations using a PHY-in-the-loop simulator built on Matlab and the Bellhop channel model~\cite{porter2011bellhop}. This simulator captures realistic UWA channel dynamics, including multipath propagation, Doppler shifts, and time-varying SNRs. 


\subsubsection{Experimental setup.} 

\begin{figure}[t!]
    \vspace{-0.8em}
    \centering
        \includegraphics[width=0.4\textwidth]{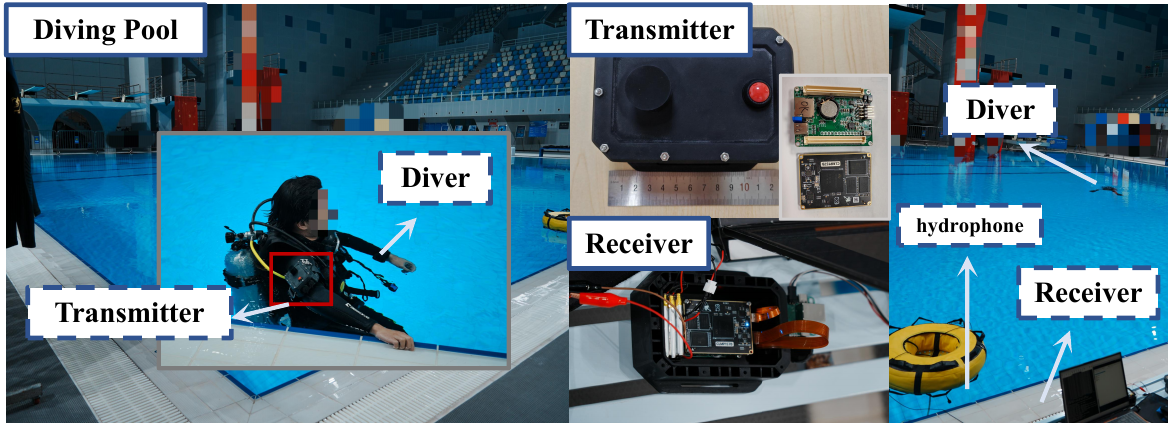}
    \caption{Diving pool environment for channel and signal profiling.}\label{fig:evir2}
    \vspace{-0.4em}
\end{figure}

\begin{figure}[htbp]
    \vspace{-0.4em}
    \centering
        \includegraphics[width=0.4\textwidth]{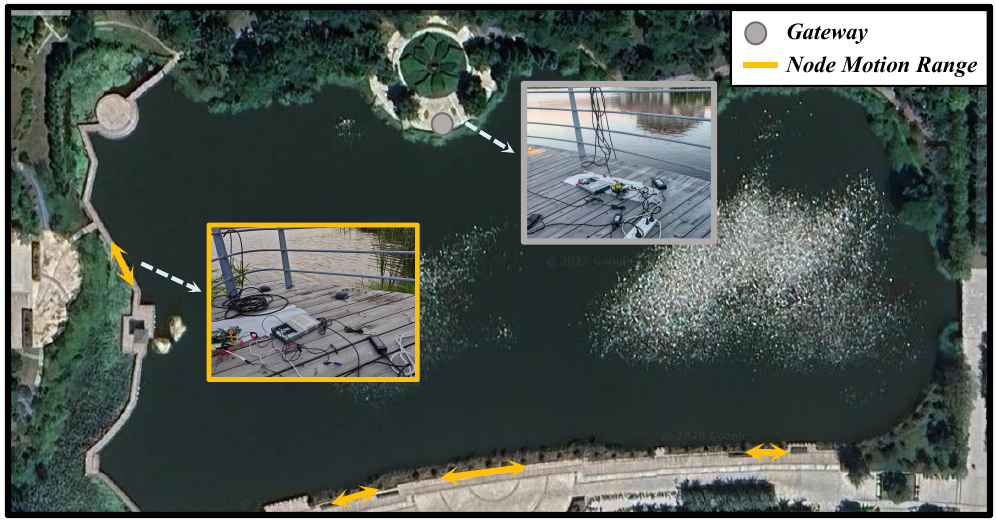}
    \caption{Lake environment for channel profiling.}\label{fig:evir1}
    \vspace{-0.8em}
\end{figure}

To gather real-world UWA channels, we conduct empirical experiments in two distinct environments: a diving pool capturing authentic diver mobility, and a lake utilizing emulated motions. \rev{Appendix~\ref{sec:channel_measurement_setup_misc} provides detailed channel measurement methodology.} 

\para{Pool Setup.} As illustrated in Fig.~\ref{fig:evir2}, our pool experiments are conducted in a $25 \times 25 \times 6$m diving pool, with acoustic nodes deployed at a depth of approximately 1.5m. To comprehensively evaluate the system under varying mobility profiles, we collect data across four distinct motion patterns relative to a stationary receiver: strictly stationary, forward-backward, left-right, and vertical (up-down) movements.


For these tests, we developed a half-duplex acoustic SDR based 
on an STM32MP157 SoC running embedded Linux. As shown in Fig.~\ref{fig:evir2}, the acoustic front-end uses a piezoelectric transducer as both transmitter and hydrophone. The platform operates at a 50kHz carrier, 20kHz bandwidth  (40--60kHz), and 200kHz AD/DA sampling rate.

\para{Lake Setup.} To validate our system in a heterogeneous environment, we conduct experiments in an artificial lake measuring $288 \times 182$m (Fig.~\ref{fig:evir1}), with devices deployed at a 2m depth. Notably, the lakebed is densely covered with decorative stones, which severely exacerbates multipath scattering.


The lake experiments use National Instruments (NI) data acquisition boards 
for continuous passband sampling. The transmitter consists of a transducer 
driven by a power amplifier, while the receiver uses a dedicated hydrophone. 
The setup operates at 120kHz sampling, 25kHz carrier, and 6kHz bandwidth 
(22--28kHz). To emulate the high-dynamic mobility of a scuba diver, the transmitter is continuously towed within a designated trajectory region (marked yellow in Fig.~\ref{fig:evir1}).



\subsection{Physical-layer Performance}\label{sec:eval:phy}


We compare \waveform against DQPSK, DSSS~\cite{pelekanakis2018adaptive}, M-ary CSS~\cite{steinmetz2022taking}, and OFDM-DQPSK~\cite{xie2009implementation}, all configured with the same symbol duration and bandwidth as \waveform. \rev{The uncoded PHY rates of \waveform\ modes and baseline waveforms are derived in Appendix~\ref{appendix:waveform_rates}.}
Among the EZCDM modes, we report MS~2, MS~3 and MS~4, corresponding to 27, 28 and 54 uncoded bits per symbol ($N=257$). All results use a single-user setting, since 1) the benchmark waveforms do not support multi-user random access; and 2) multi-user cross-root interference is effectively whitened, this remains representative of concurrent-access performance.




\begin{figure}[tbp]
    \centering
    \subfloat[$\xi_{CIR}^{*\text{mean}}\ge 0.25$ (low/moderate variability)]{%
        \includegraphics[width=0.40\textwidth]{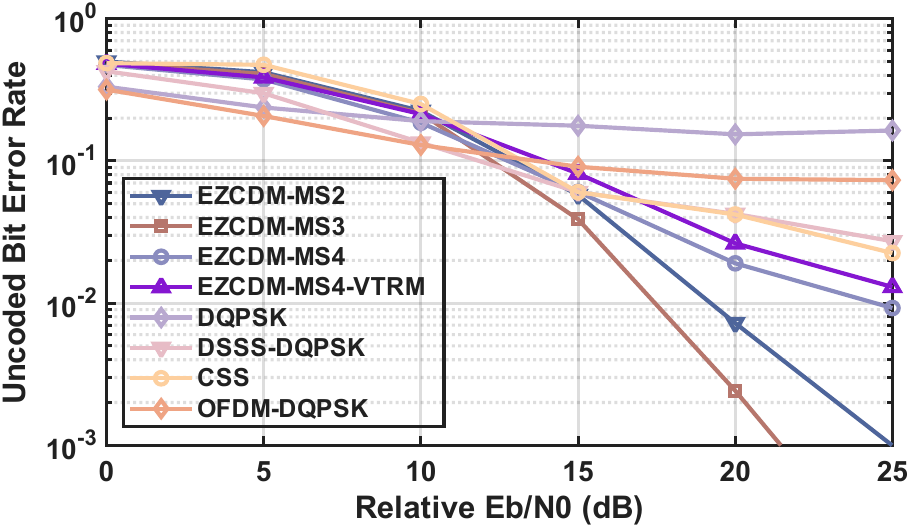}
    }\hfill
    \subfloat[$\xi_{CIR}^{*\text{mean}}<0.25$ (high variability)]{%
        \vspace{-1.2em}
        \includegraphics[width=0.40\textwidth]{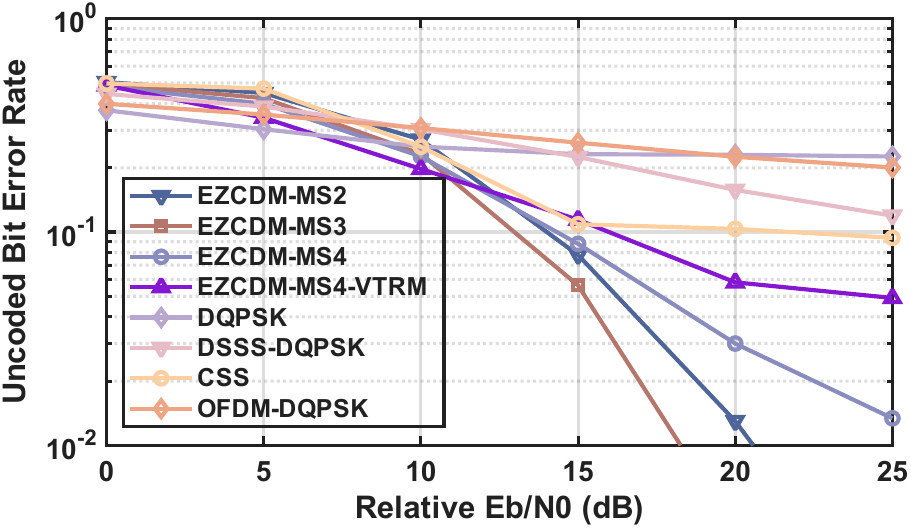}
    }%
    \caption{Channel-trace-driven evaluation: uncoded BER vs.\ $E_b/N_0$ of \waveform\ (MS~2, MS~3, MS~4, MS4-VTRM), DSSS, CSS, and OFDM.}
    \label{fig:comparison}
    \vspace{-0.8em}
\end{figure}

\subsubsection{Real-world channel-based evaluation.} 

We use both the measured channels and the open-source channels obtained from the Watermark dataset~\cite{van2017watermark}. We refer to channels with $\xi_{CIR}^{*\text{mean}}>0.5$ are defined as low variability channels, those with $0.25<\xi_{CIR}^{*\text{mean}}\leq0.5$ as moderate variability channels, and channels with $\xi_{CIR}^{*\text{mean}}\leq0.25$ are referred to as high variability channels.
The replay filter method is adopted to apply real-world channel responses to different waveforms, incorporating both instantaneous Doppler shifts and multipath effects. 


We consider uncoded BER as a function of $E_b/N_0$, which normalizes the energy differences caused by the distinct numbers of information bits carried per waveform block. Because the replay-filter method replays CIRs extracted from real-world measurements, the recorded signals already contain residual measurement noise; the $E_b/N_0$ values shown here therefore represent the ratio of the added signal energy to this existing noise floor, rather than an absolute noise-free reference.
As shown in Figs.~\ref{fig:comparison}(a)(b), \waveform outperforms the baseline waveforms under both channel conditions, with the advantage being more pronounced in highly dynamic channels.
Under $\xi_{CIR}^{*\text{mean}}\ge 0.25$, MS~3 achieves orders-of-magnitude lower BER than other waveforms at relative $E_b/N_0\ge 20$\,dB.
At a BER of $0.02$, EZCDM achieves at least a $5$\,dB $E_b/N_0$ gain over CSS, the best-performing baseline.
Under $\xi_{CIR}^{*\text{mean}}<0.25$, the gain increases to $10$\,dB at a BER of $0.01$, confirming \waveform's robustness under severe channel dynamics.


We also compare MS~4 against a virtual time-reversal mirror (VTRM) baseline, a recent non-coherent demodulator that correlates the received signal with the conjugate of the preamble and takes the peak of the resulting profile---without exploiting the multipath structure as our detector does. MS~4 consistently outperforms this VTRM counterpart across both channel regimes, confirming the benefit of multipath-weighted combining over single-peak detection.

\subsubsection{Real-world signal-based evaluation.}
\begin{figure}[tbp]
    \centering
         \includegraphics[width=0.40\textwidth]{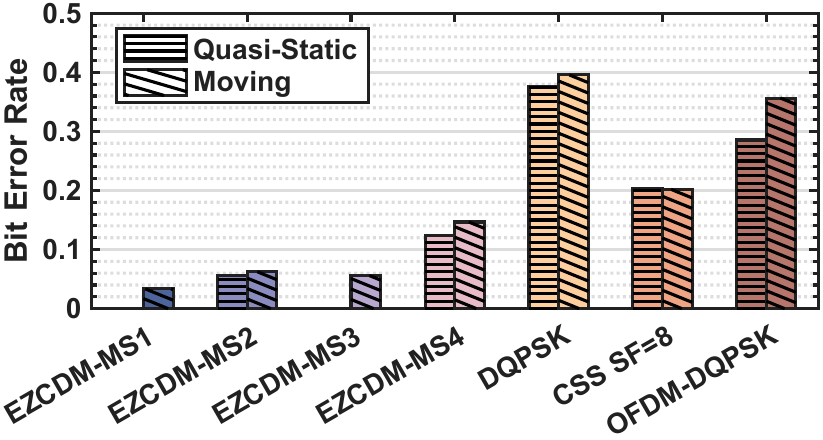}
    \vspace{-1em}
    \caption{Signal-trace-driven evaluation: BER of \waveform\ and baseline waveforms decoded from recorded pool-environment passband signals.}\label{fig:rw_data}
    \vspace{-0.8em}
\end{figure}

As a complementary signal-driven validation, we transmit \waveform and the benchmark waveforms through the pool testbed and directly decode the recorded passband signals without channel estimation or equalization. We use 1/2 LDPC coding for all waveforms, and the per-sample SNR is set to around 8~dB.
Fig.~\ref{fig:rw_data} reports the per-mode BER under the four representative motion patterns whose CIRs are shown in Fig.~\ref{fig_channel}(a)--(d).
In the quasi-static regime (Fig.~\ref{fig_channel}(a)), MS~1 and MS~2 achieve near-error-free transmission, whereas all baselines exhibit BER above 0.2.
Under highly dynamic conditions (Fig.~\ref{fig_channel}(b)--(d)), the BER of MS~1 and MS~3 outperforms other waveforms by approximately an order of magnitude.
Although MS~2 and MS~4 suffer degraded performance in the most severe dynamic scenarios, 42\% and 63\% of their packets, respectively, are decoded without error---far exceeding the sub-10\% error-free rate of all baseline waveforms.

\subsection{Power Control Performance}\label{sec:eval:power}


To validate the closed-loop power control mechanism described in \S\ref{sec:power_control}, we conduct a two-user experiment in the pool environment using the developed prototype (see Fig.~\ref{fig:evir2}).
The near user is placed close to the gateway (2m) while the far user is positioned at the pool edge (6m), creating an initial received-power imbalance of approximately~6~dB. $\delta_p$ is set to 25\% of the target power, and $\Delta_p$ is set to 10\% of the maximal gain.
Across consecutive update rounds, the normalized preamble-correlation magnitudes of both users converge: after 5 rounds, both reach within~0.14~dB of the target level, and the residual inter-user power imbalance is reduced to less than~0.27~dB (averaged over ten repeated trials).
These results confirm that the proposed closed-loop mechanism can effectively mitigate near--far imbalance in a practical deployment.

\subsection{Network-layer Performance}



\para{Setup.}
We simulated a 12-node star-topology network (1 gateway, 11 users) 
uniformly deployed within a $70$\,m radius (corresponding to a 47\,ms 
propagation delay). The system bandwidth is $20$\,kHz and the carrier 
frequency is $50$\,kHz. The frame/slot duration is fixed at $T = 378$\,ms, 
consisting of a 23-symbol packet (3 preamble symbols, $N=257$ samples per 
symbol, $10/11$ CP ratio, 325\,ms) and a 53\,ms guard interval that 
accommodates the maximum propagation delay plus the timing 
uncertainties of the proposed MAC protocol ($6$\,ms).

We use a PHY-in-the-loop simulator built on Matlab and the Bellhop channel 
model~\cite{porter2011bellhop}. Matlab manages MAC logic and passes each 
transmission to an external PHY block, which maps signals to Bellhop 
channels, performs waveform-specific synchronization, soft demodulation, 
LDPC decoding, and CRC verification. Channel dynamics follow a latent 
Ornstein-Uhlenbeck (OU) process augmented with a compound Poisson jump 
component, with time-varying intensity coupled to $\xi_{CIR}^{*\text{mean}}$; the measured SINR variation is $[-13,\text{dB}, 10,\text{dB}]$.
Packets at each node are generated according to a Poisson process with 
mean arrival rate $\lambda \in [0.04, 0.22]$~packets/slot. Each simulation 
run lasts over $1000$\,s, and results are averaged over 10 independent runs.

The following protocols are considered:

\para{$\rhd$ \system-(CLA/ELA).} 
The proposed conservative LA (CLA, Sec.~\ref{sec:link_adaptation}) selects 
a common MS from the precomputed MS--SINR--CIR-Similarity--PER LUT. For 
comparison, an exhaustive LA benchmark (ELA) iterates through all available modes after each round and chooses the one that maximized goodput on the just-observed channel; this mode is then used for the next round, which is equivalent to having perfect knowledge of the last-round channel state to select the MS for the upcoming transmission. Note that the ELA scheme is infeasible in practice due to the exhaustive search and perfect channel knowledge.

\para{$\rhd$ S-ALOHA-EZCDM.} 
Each user independently transmits with probability inversely 
proportional to the number of users per round; overlapping packets are 
treated as a collision, and the CLA scheme selects the MS.

\para{$\rhd$ TDMA-EZCDM.} 
Each user is assigned a fixed time slot and transmits with the EZCDM 
waveform, using the same CLA look-up based on last-round channel information.

\para{$\rhd$ CDMA and ZCMod.} 
Users transmit in each slot. Both use the same distinct-sequence per-user as \waveform, differing only in the demodulation stage: CDMA uses peak-based matching, while 
ZCMod employs single-shift template matching~\cite{zhang2025high}.



\begin{figure}[t!]
    \centering
        \includegraphics[width=0.40\textwidth]{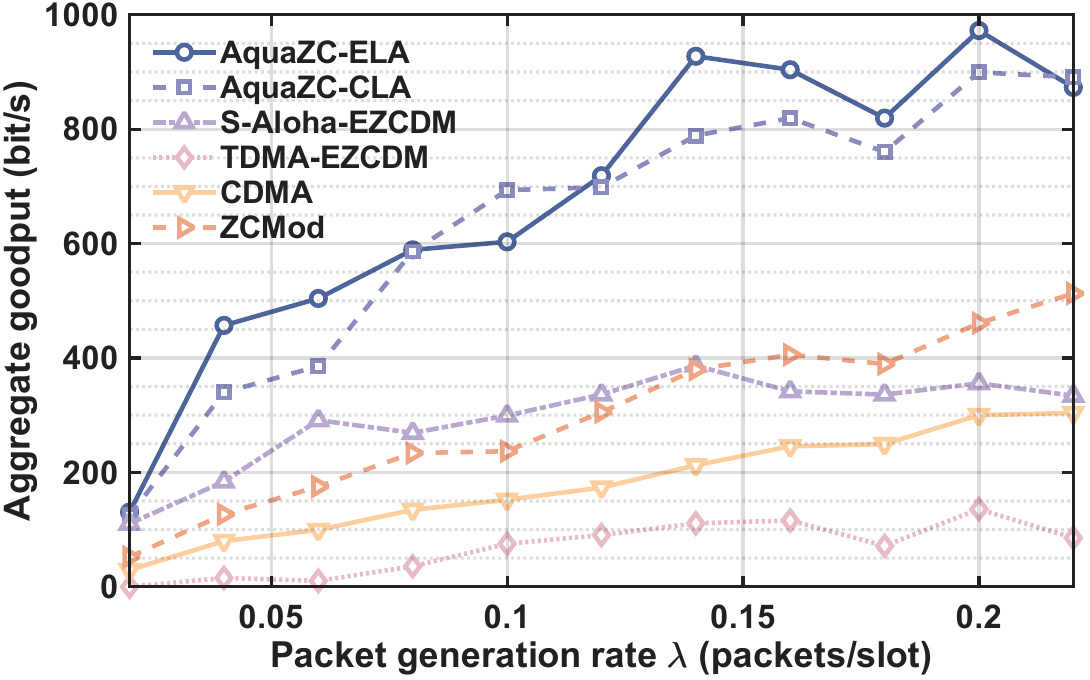}
    \vspace{-1em}
    \caption{Network throughput of different MACs.}
    \label{fig:fig_tpt}
    \vspace{-1em}
\end{figure}

\para{Results.}
As shown in Fig.~\ref{fig:fig_tpt}, \system-CLA achieves substantial throughput gains across the entire spectrum of packet generation rates. At a workload of 0.16 packets/slot, it delivers improvements of 7.13$\times$, 3.34$\times$, 2.40$\times$,  and 2.02$\times$ over TDMA-EZCDM, CDMA, S-ALOHA-EZCDM and ZCMod respectively, while the peak throughput is enhanced by 10.6$\times$, 4.56$\times$, 2.71$\times$, and 2.92$\times$, respectively.

We analyze the gains case by case:

\para{$\rhd$ vs.~TDMA-EZCDM.} \system benefits from both concurrent decoding and on-demand access that avoids idle slots; TDMA could in principle adopt higher-rate OFDM, but OFDM performs poor in dynamic channel (\S\ref{sec:eval:phy}).

\para{$\rhd$ vs.~S-ALOHA-EZCDM.} Holding the waveform and LA fixed, \system mainly benefits from concurrent decoding.

\para{$\rhd$ vs.~CDMA and ZCMod.} All use distinct-sequence concurrent transmission, so the gain is from the waveform design: EZCDM's higher uncoded rate and intra-symbol differential receiver deliver more decodable bits per concurrent symbol.

\para{$\rhd$ vs.~\system-ELA.} Holding the waveform and MAC fixed, CLA and ELA exhibit comparable performance while ELA is infeasible in practice, confirming that the conservative strategy is effective under different packet rates. 

\rev{
\section{Related Work} \label{sec:relatework}

\textbf{Underwater acoustic communication and networking.}
Single-carrier modulations (M-FSK~\cite{wax1981mfsk}, M-PSK~\cite{stojanovic1994phase}, CSS~\cite{lei2012implementation,jia2022two,steinmetz2022taking,petroni2023feasibility}, DSSS~\cite{pelekanakis2018adaptive}, CDMA~\cite{yang2015spatially}) have been widely studied alongside multi-carrier schemes (OFDM~\cite{li2008multicarrier}, OTFS~\cite{BITS2022}, OCDM~\cite{ouyang2016orthogonal}); more recently, smartphone- and smartwatch-based UAC systems have been explored for underwatermessaging and SOS signal detection~\cite{chen2022underwater,yang2023aquahelper,yang2024nerual}.
On the MAC side, common designs include random access~\cite{chirdchoo2007aloha}, handshake-based~\cite{molins2006slotted}, and TDMA-based~\cite{kredo2009stump} protocols.
Both layers share the same limitation: conventional PHY waveforms either offer low rates or require accurate channel estimation, while MAC protocols treat overlapping transmissions as destructive collisions. By designing a waveform that achieves both high rate and concurrent-access capability, our work redefines the role of the MAC layer.

\textbf{Link adaptation in UAN.} Prior LA schemes tune modulation based on post-equalization SNR in point-to-point links~\cite{mani2008adaptive,wan2014adaptive} or leverage sparse channel estimation and pre-trained classifiers~\cite{radosevic2013adaptive,huang2020adaptive}. In contrast, our LA jointly adapts the differential alphabet and $\tau$ based on both CIR similarity and worst-case SINR in a multi-user concurrent-access setting.
}

\rev{
\section{Conclusion}    \label{sec:conclusion}

This paper presented \system, a high-rate cross-layer concurrent-access system that combines EZCDM---an equidistant ZC division multiplexing waveform with a lightweight intra-symbol differential receiver---with root-separated asynchronous uplinks, closed-loop power control, and overlap- and CIR-aware link adaptation. Experimental evaluations using custom SDR prototypes, replay of measured and open-source UWA channels, and PHY-in-the-loop network simulations confirm that \system\ substantially outperforms both non-coherent and coherent baseline waveforms as well as conventional MAC protocols. Future work includes real-time SoC implementation and channel access control that adapts to traffic load and multi-user reception capability. 
}

\appendices

\rev{
\section{Validation of Whitened Cross-User Interference}\label{sec:whitevali}
ZC sequences with distinct roots exhibit bounded and consistently low cyclic cross-correlation (Sec.~\ref{sec:zcseq}). In a multi-user random-access setting, assigning a different root to each user transforms concurrent access into an equivalent single-user problem---each receiver sees the target signal plus residual cross-root interference that
superimposes as an elevated noise floor. Section~\ref{sec:amc} exploits this property to treat multi-user interference as additive white noise.
This appendix provides the quantitative justification: we measure the spectral flatness of the residual cross-root interference under asynchronous random access and compare it directly against the AWGN benchmark.

We examine three interference regimes through simulation: (i) \emph{Full Concurrent}: all users' frame starts align within the CP window, producing complete temporal overlap of the ZC-shift-domain blocks; (ii) \emph{Partial Concurrent}: users' arrival times are independently randomized according to the beacon-framed access in Sec.~\ref{sec:amc}, creating arbitrary partial overlaps; and (iii) AWGN as the reference. The goal is to statistically compare the spectral characteristics of the first two regimes against the white-noise benchmark.

Recall from Eq.~\eqref{eq:demod} that the ZC-shift-domain spectrum $\mathbf{z}$ is the matched cyclic-correlation output. Its absolute value $|\mathbf{z}|$, which we denote by the vector $\bm{r}$, is proportional to the cyclic correlation magnitude spectrum \cite{gardner1994cyclostationarity}. In the absence of a signal, $\bm{r}$ reduces to the post-correlation noise spectrum. To quantify spectrum whiteness, we introduce the spectral Flatness Measure (SFM) \cite{dubnov2004generalization,peeters2004audio}, a compact statistical descriptor defined as
\begin{equation}
\mathrm{SFM} = \frac{\exp\left( \frac{1}{N} \sum_{i=1}^{N} \ln r[i] \right)}{\frac{1}{N} \sum_{i=1}^{N} r[i]}.
\end{equation}
By construction, $\mathrm{SFM}\in[0,1]$: a value close to~1 indicates a perfectly flat (white) spectrum, whereas a value close to~0 indicates a highly structured, tonally concentrated spectrum.

This validation normalizes the interference and focuses on the intrinsic distribution pattern, independent of signal magnitude. Since each frame is treated as statistically independent, local anomalies are preserved rather than smoothed out, ensuring a faithful characterization of spectrum behavior.

Fig.~\ref{fig:muint}(a) illustrates examples of cyclic correlation under different levels of concurrency. As the number of concurrent users increases, the background noise floor rises systematically, but the overall spectral envelope remains unchanged---the shape is preserved, only the level shifts. This separation of level and shape is the signature of whitened interference.
Fig.~\ref{fig:muint}(b) shows that both Full Concurrent and Partial Concurrent interference scenarios yield SFM values around 0.85, closely aligning with those of AWGN. Quantitatively, the mean SFM difference between either concurrent regime and AWGN is below~0.05 across the tested concurrency levels, confirming a strong tendency toward spectral whiteness in both interference types.

In summary, this validation establishes that multi-user ZC cross-root interference under asynchronous access is effectively whitened. This property provides the analytical foundation for two simplifications used throughout the paper: (i) the worst-case effective SINR in Eq.~\eqref{eq:worst_case_sinr} needs only power-level information, not phase or structural interference models; and (ii) single-user PER results obtained through replay or Bellhop channels remain representative of multi-user performance in system-level simulation.



\begin{figure}[tbp]
    \centering
    \subfloat[]{%
        \includegraphics[width=0.233\textwidth]{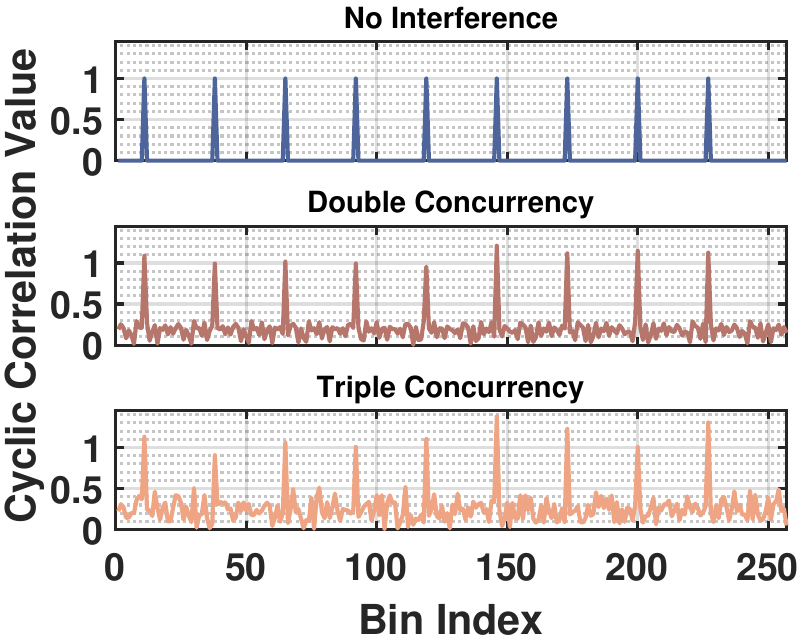}%
    }\hfill
    \subfloat[]{%
        \includegraphics[width=0.233\textwidth]{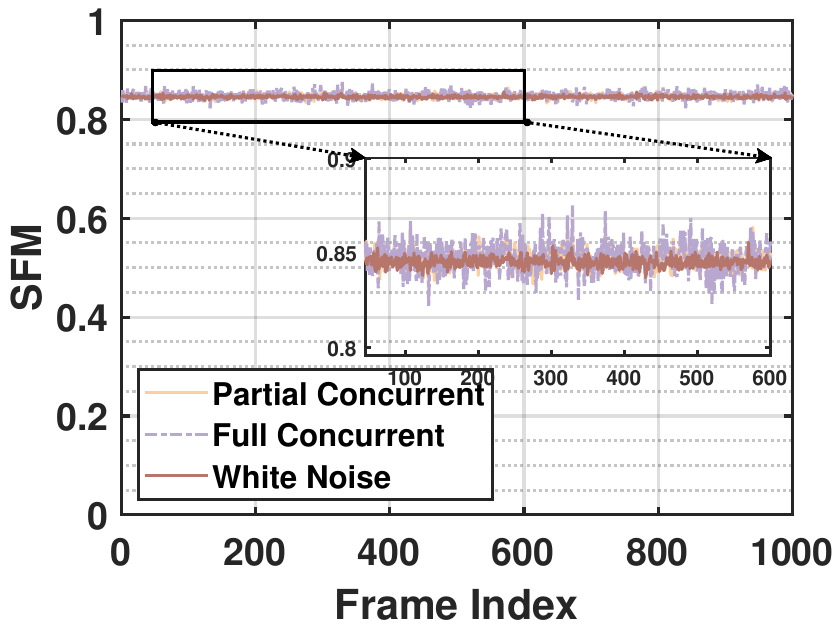}
        }%

    \caption{(a) Influence of equal-power multi-user interference on cyclic correlation spectrum; (b) SFM of cyclic correlation spectra under different types of interference.}
    \label{fig:muint}
\end{figure}

}

\rev{
\section{CP Window-Offset Robustness}\label{appendix:cp_window}
The CP model in Sec.~\ref{subsec:ezcdm} assumes that the receiver retains an
$N$-sample window after timing alignment. This appendix makes explicit how
much residual window displacement that model tolerates. Using the notation of
Eq.~\eqref{eq:mod}, the CP-extended version of $\mathbf{u}$ can be written as
\begin{equation}
    u_{\mathrm{cp}}[n]=u\!\left([n]_N\right),
    \qquad n=-N_{\mathrm{cp}},\ldots,N-1,
    \label{eq:cp_extension}
\end{equation}
where $[\cdot]_N$ denotes reduction modulo $N$. For the channel in
Sec.~\ref{subsec:ezcdm}, whose nonzero taps lie at delays $0,\ldots,D_h$, let
the receive window begin $\Delta$ samples earlier than the nominal block
boundary. Its $n$-th sample is
\begin{equation}
    y_{\Delta}[n]
    =\sum_{d=0}^{D_h}h[d]u_{\mathrm{cp}}[n-\Delta-d]
     +v_{\Delta}[n],
    \quad n=0,\ldots,N-1.
    \label{eq:cp_offset_samples}
\end{equation}
If
\begin{equation}
    0\leq\Delta\leq N_{\mathrm{cp}}-D_h,
    \label{eq:cp_offset_range}
\end{equation}
all required samples remain inside the CP-protected interval, and
Eq.~\eqref{eq:cp_offset_samples} is still a circular convolution. Define the
length-$N$ circular-shift operator $\mathbf{S}_{\Delta}$ by
$[\mathbf{S}_{\Delta}\mathbf{a}]_n=a([n-\Delta]_N)$. The retained block and
its ZC-shift-domain spectrum are then
\begin{align}
    \mathbf{y}_{\Delta}
    &=\mathbf{S}_{\Delta}\mathbf{H}_{c}\mathbf{u}
      +\mathbf{v}_{\Delta}, \notag\\
    \mathbf{z}_{\Delta}
    &=\frac{1}{N}\mathbf{C}_{r}^{H}\mathbf{y}_{\Delta}
      =\mathbf{S}_{\Delta}\mathbf{H}_{c}\mathbf{x}
      +\mathbf{v}_{z,\Delta}.
    \label{eq:cp_offset_shift_domain}
\end{align}
The second equality follows because $\mathbf{S}_{\Delta}$,
$\mathbf{H}_{c}$, and $\mathbf{C}_{r}$ are circulant and commute, while
$\mathbf{C}_{r}^{H}\mathbf{C}_{r}=N\mathbf{I}_N$. Hence, an admissible
window displacement cyclically translates the entire channel-broadened
ZC-shift profile; it does not change its shape or the relative spacing among
the activated coefficients. The front end can absorb this common translation
into the estimated path offsets. This is a bounded robustness result rather
than a claim that an arbitrary timing error is harmless: displacement beyond
Eq.~\eqref{eq:cp_offset_range} can introduce samples from an adjacent block
and invalidate the circular model.

Fig.~\ref{fig:cpzp_diff}(a) illustrates this behavior for a five-sample
receive-window displacement: the CP case preserves the ZC-shift-domain
profile up to a common cyclic translation. With zero padding (ZP), an early
window instead contains zeros while omitting samples from the other end of
the block. The resulting truncation cannot be represented by a circular-shift
operator and causes leakage across the ZC-shift bins, as shown in
Fig.~\ref{fig:cpzp_diff}(b). This preservation of the receiver model is why
we use CP rather than ZP.

\begin{figure}[t!]
    \centering
    \subfloat[CP padding]{%
        \includegraphics[width=0.235\textwidth]{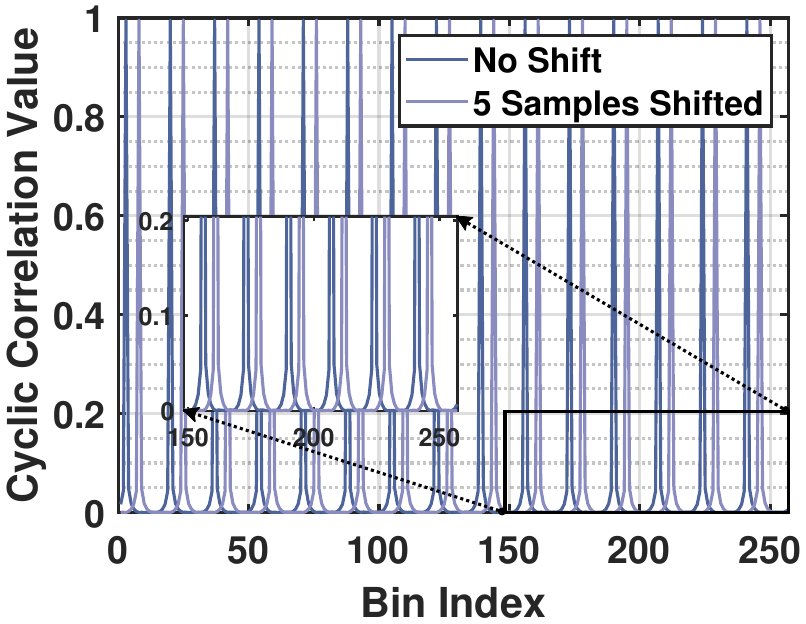}%
    }\hfill
    \subfloat[Zero padding]{%
        \includegraphics[width=0.235\textwidth]{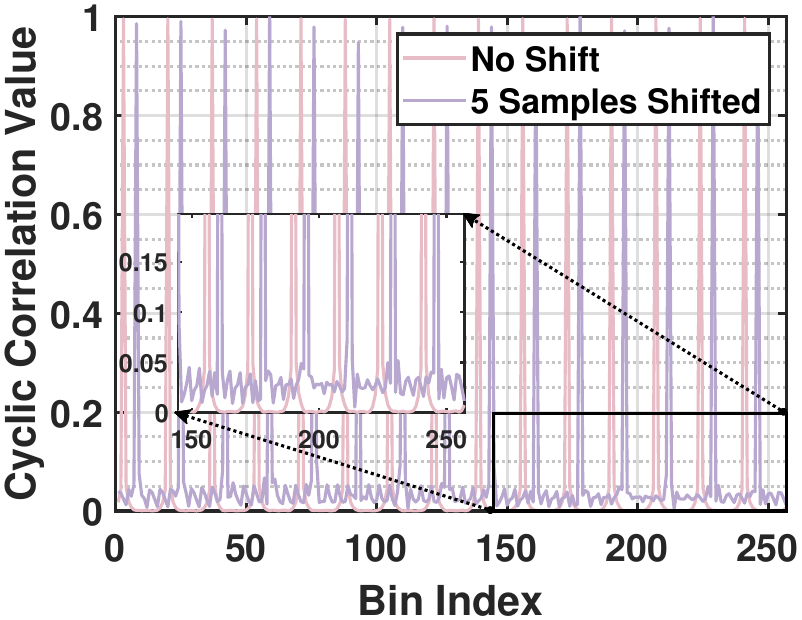}%
    }
\caption{Effect of a five-sample receive-window displacement on the
    ZC-shift-domain spectrum with (a) cyclic-prefix padding and (b)
    zero padding.}
    \label{fig:cpzp_diff}
\end{figure}

}

\rev{
\section{LLR Conversion for Channel Decoding} \label{appendix:llr}
In the system-level simulator, the LDPC decoder consumes bit log-likelihood
ratios (LLRs) derived from the candidate metrics in Eq.~\eqref{ori_p}. The
folded path profile and its weights $\{w_d\}$ are shared across the block, but
the metric $\Lambda_m[\ell]$ is computed separately for every differential
index $\ell$ by combining only its observations across the selected path
offsets $d\in\widehat{\mathcal{D}}$; metrics from different values of $\ell$
are not combined. For the $B$-bit binary label of candidate $m$, let
$b_p(m)\in\{0,1\}$ denote its $p$-th bit, where $p=0,\ldots,B-1$. We first
convert each candidate metric into a normalized symbol probability,
\begin{equation}
    P_m[\ell]=
    \frac{\exp\!\left(\Lambda_m[\ell]/T_{\Lambda}\right)}
    {\sum_{n=0}^{M-1}\exp\!\left(\Lambda_n[\ell]/T_{\Lambda}\right)},
    \label{eq:metric_to_probability}
\end{equation}
for $m=0,\ldots,M-1$ and $\ell=1,\ldots,K-1$. Here,
$T_{\Lambda}>0$ is the likelihood-scale parameter, and we currently set
$T_{\Lambda}=1$. 

The bit LLR supplied to the LDPC decoder is then
\begin{align}
    \mathcal{L}_{p}[\ell]
    &=\log
    \frac{\sum_{m:b_p(m)=0}P_m[\ell]}
         {\sum_{m:b_p(m)=1}P_m[\ell]} \notag\\
    &=\log
    \frac{\sum_{m:b_p(m)=0}
    \exp(\Lambda_m[\ell]/T_{\Lambda})}
         {\sum_{m:b_p(m)=1}
    \exp(\Lambda_m[\ell]/T_{\Lambda})}.
    \label{eq:ezcdm_llr}
\end{align}
Thus, one waveform block produces $(K-1)B$ LLRs. In implementation, the two
log-sum-exp terms in Eq.~\eqref{eq:ezcdm_llr} are evaluated with the standard
max-subtraction stabilization to avoid overflow and underflow.
}

\rev{
\section{LUT for System-level Simulation}\label{appendix:lut}

We build the look-up table off-line as follows. First, we assemble a
channel set spanning the CIR-similarity range observed in our measurements
(Sec.~\ref{sec:eval:exp}). For each target similarity interval,
Bellhop~\cite{porter2011bellhop} is used to generate independent UWA channel
realizations with the corresponding multipath dynamics. Each candidate MS is
then transmitted through randomly selected channels from this set. At each
(SINR, CIR-similarity) bin, signal-level decoding trials produce the CRC
packet-error rate $\widehat{\mathrm{PER}}(m,\gamma,\xi)$.
Fig.~\ref{fig:amc_lut_full} shows the resulting
MS--SINR--CIR-Similarity--PER surfaces. Their staircase transitions reflect
the distinct robustness--rate operating regions of the four MSs.

At run time, the selector first requires a predicted PER no larger than 0.1
for every valid observed link. Among those feasible modes, it maximizes the
worst-link predicted goodput
$R_m[1-\widehat{\mathrm{PER}}(m,\gamma,\xi)]$, as defined in
Eq.~\eqref{eq:common_ms_selection}. If no mode satisfies the PER target, it
falls back to the mode with the smallest worst-link predicted PER.

\begin{figure}[tbp]
    \centering
    \subfloat[PER performance in UWA channels with $\xi_{CIR}^{*\text{mean}}>0.5$ (low variability)]{%
        \includegraphics[width=0.48\textwidth]{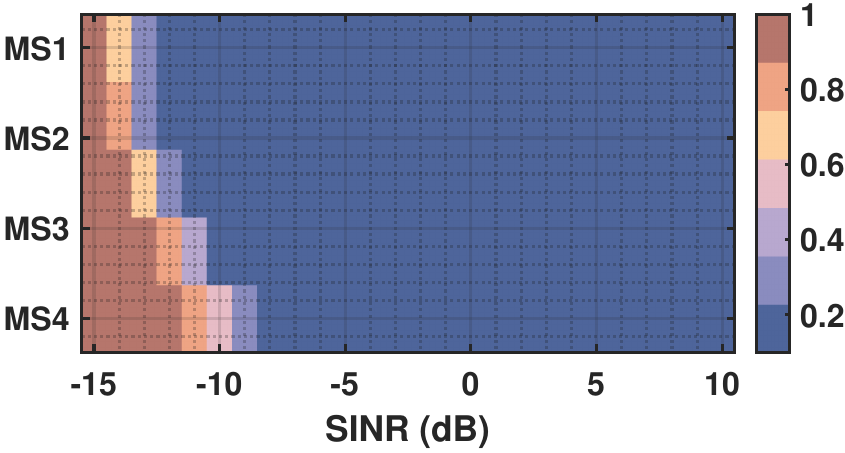}%
    }
    \hfill
    \subfloat[PER performance in UWA channels with $0.25<\xi_{CIR}^{*\text{mean}}\leq0.5$ (moderate variability)]{%
        \includegraphics[width=0.48\textwidth]{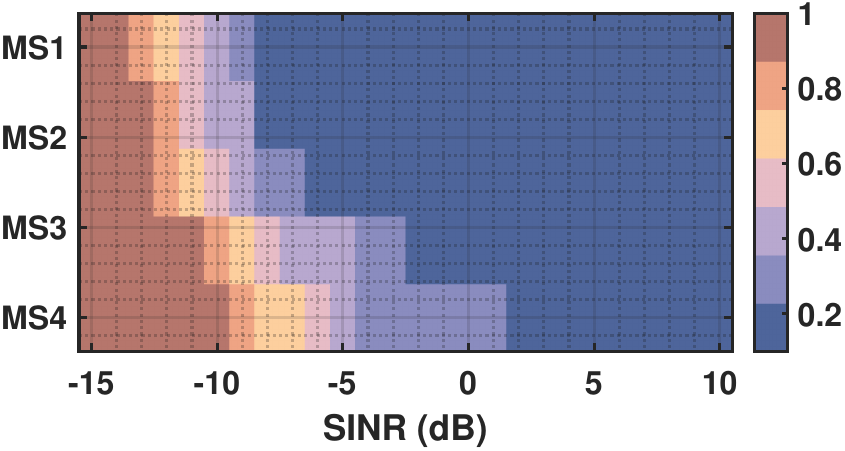}
        }%
        \hfill
    \subfloat[PER performance in UWA channels with $\xi_{CIR}^{*\text{mean}}\leq0.25$ (high variability)]{%
        \includegraphics[width=0.48\textwidth]{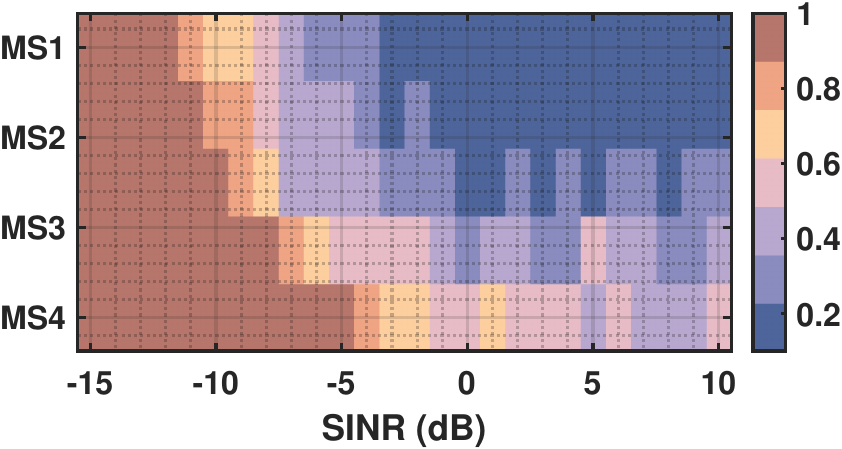}
        }%

    \caption{MS--SINR--CIR-Similarity--PER look-up table under UWA channels with varying CIR similarity.}
    \label{fig:amc_lut_full}
\end{figure}

}

\rev{
\section{Channel Measurement Methodology} \label{sec:channel_measurement_setup_misc}

Across both the pool and lake environments (see Section~\ref{sec:eval:exp} for the hardware and deployment details), we adopt a standardized sounding approach: a linear chirp followed by a pseudo-random binary sequence (PRBS) serves as the probe frame. The receiver uses the chirp for coarse packet synchronization and then performs cross-correlation with the PRBS to extract the Channel Impulse Response (CIR) at high resolution.

Although utilizing the identical pool environment and SDR platform, the background measurements were exclusively designed for \textit{channel sounding} to profile UWA dynamics, rather than transmitting \waveform communication payloads. Specifically, the transmitter generated a specialized probe frame operating at a 50\,kHz carrier frequency with a 200\,kHz sampling rate. The frame consists of two parts:
1) A 0.3-second Linear Frequency Modulation (LFM) chirp used for precise frame synchronization.
2) A continuous pseudo-random binary sequence (PRBS) for channel probing. The PRBS is generated using a 511-length m-sequence with a chip duration of 0.05\,ms (yielding a 20\,kHz bandwidth) and is modulated via BPSK.

At the stationary receiver, the offline processing first utilizes cross-correlation with the reference chirp to locate the temporal start of the frame. Subsequently, a sliding matched filter correlates the received passband signal with the local PRBS replica across the delay domain. This process precisely extracts the high-resolution, time-varying delay profiles ($h(t,\tau)$). 

During these sounding sessions, the transmitting node was maneuvered through specific kinematic patterns---hovering, horizontal motion (at approximately 0.5\,m/s), and vertical displacement. These isolated empirical measurements accurately capture the severe Doppler scaling and multipath evolution characteristic of diver mobility, which directly motivates the core physical-layer innovations of \system.
}

\rev{
\section{PHY Rates of Different Waveforms}
\label{appendix:waveform_rates}

This appendix derives the uncoded data-only PHY rates of \waveform\ and the baseline waveforms used in the physical-layer evaluation (\S\ref{sec:eval:phy}). Rates include the cyclic-prefix (CP) duration but exclude packet-level overhead (preamble, midamble, postamble, and guard intervals). All waveforms are assumed to operate at the same bandwidth $W$ and the same symbol duration $T_{\mathrm{sym}}$ for fair comparison.

\para{EZCDM Modes.}
For mode $m$ with ZC length $N=257$, equidistant spacing $\tau_m$, and differential alphabet size $M_m$, the number of active shift dimensions is $K_m=\lfloor N/\tau_m\rfloor$. The first active dimension serves as a differential reference, so each EZCDM symbol carries $L_m=(K_m-1)\log_2 M_m$ uncoded payload bits. With CP ratio $\rho=N_{\mathrm{cp}}/N$, the symbol duration is $T_{\mathrm{sym}}=N(1+\rho)/W$. The uncoded PHY rate is therefore
\begin{equation}
    R_m^{\mathrm{EZCDM}}=\frac{L_m}{T_{\mathrm{sym}}}=\frac{(K_m-1)\log_2 M_m}{N(1+\rho)}\,W.
    \label{eq:ezcdm_rate}
\end{equation}
Table~\ref{tab:all_rates} summarizes the resulting per-mode payloads and normalized rates.

\para{Baseline Waveforms.}
\textit{DSSS-BPSK} uses a 256-chip spreading sequence with BPSK modulation, delivering 1~uncoded bit per symbol: $R_{\mathrm{DSSS}}=1/T_{\mathrm{sym}}=W/[N(1+\rho)]$.
\textit{CSS} ($SF=8$) uses chirp spreading with $2^{SF}=256$ samples per chirp, delivering $SF=8$~uncoded bits per symbol: $R_{\mathrm{CSS}}=8/T_{\mathrm{sym}}=8W/[N(1+\rho)]$.
\textit{OFDM-DQPSK} uses 256 subcarriers with differential QPSK (2~bits per subcarrier), delivering $256\times2=512$~uncoded bits per OFDM symbol: $R_{\mathrm{OFDM}}=512/T_{\mathrm{sym}}=512W/[N(1+\rho)]$.
\textit{DQPSK} is a single-carrier differential QPSK baseline carrying 2~bits per symbol: $R_{\mathrm{DQPSK}}=2/T_{\mathrm{sym}}=2W/[N(1+\rho)]$.
\textit{VTRM} uses the same EZCDM mode-4 parameters ($\tau=9$, DQPSK) and thus has the same payload (54~bits/symbol) and rate as MS~4, differing only in the demodulation algorithm.

\begin{table}[tbp]
\centering
\caption{Uncoded PHY rates of \waveform\ modes and baseline waveforms. All rates include the CP ($\rho=N_{\mathrm{cp}}/N$) and assume identical bandwidth $W$ and symbol duration $T_{\mathrm{sym}}=N(1+\rho)/W$, with $N=257$.}
\label{tab:all_rates}
\scriptsize
\begin{tabular}{l c c c}
\toprule
\textbf{Waveform} & \textbf{bits/sym.} & \textbf{$R/W$ (bits/s/Hz)} & \textbf{Rel.~to MS~1} \\
\midrule
\multicolumn{4}{c}{\textit{\waveform\ modes}} \\
\cmidrule(lr){1-4}
MS~1 (DBPSK, $\tau{=}17$)  & 14  & $14/[257(1+\rho)]$  & $1.000\times$ \\
MS~2 (DBPSK, $\tau{=}9$)   & 27  & $27/[257(1+\rho)]$  & $1.929\times$ \\
MS~3 (DQPSK, $\tau{=}17$)  & 28  & $28/[257(1+\rho)]$  & $2.000\times$ \\
MS~4 (DQPSK, $\tau{=}9$)   & 54  & $54/[257(1+\rho)]$  & $3.857\times$ \\
\midrule
\multicolumn{4}{c}{\textit{Baseline waveforms}} \\
\cmidrule(lr){1-4}
DQPSK            & 2   & $2/[257(1+\rho)]$    & $0.143\times$ \\
DSSS-BPSK        & 1   & $1/[257(1+\rho)]$    & $0.071\times$ \\
CSS ($SF{=}8$)   & 8   & $8/[257(1+\rho)]$    & $0.571\times$ \\
OFDM-DQPSK       & 512 & $512/[257(1+\rho)]$  & $36.571\times$ \\
\bottomrule
\end{tabular}
\end{table}

For the EZCDM family, the rate spans a factor of $3.86\times$ from MS~1 to MS~4. MS~4 carries twice the payload of MS~2 ($54$ vs.\ $27$~bits/symbol) at the same spacing $\tau=9$, isolating the effect of the differential alphabet size. MS~2 and MS~3 have nearly equal payloads ($27$ vs.\ $28$~bits/symbol) but employ different spacing--alphabet combinations, enabling approximately rate-matched comparisons. Among the baselines, OFDM-DQPSK achieves the highest nominal rate ($512$~bits/symbol) because all subcarriers carry independent payload, but its BER performance degrades substantially under time-varying UWA channels without accurate channel estimation (\S\ref{sec:eval:phy}).
}

\rev{
\section{Notation}
\label{appendix:notation}

\begin{table}[htbp]
\centering
\caption{Summary of key notation.}
\label{tab:notation}
\scriptsize
\begin{tabular}{l l}
\toprule
\multicolumn{2}{c}{\textbf{System Model (Sec.~\ref{sec:bg})}} \\
\midrule
$f_c$                & Carrier frequency \\
$h(t,\tau)$          & Time-varying baseband channel impulse response (CIR) \\
$a_p = v_p/c$        & Path-dependent Doppler scale \\
$x(t)$               & Complex-baseband packet transmit waveform \\
$y(t)$               & Received baseband signal \\
$\nu^{\mathrm{cfo}}$ & Carrier-frequency offset (CFO) \\
$\mathcal{U}(t)$     & Set of active users at time $t$ \\
\midrule
\multicolumn{2}{c}{\textbf{PHY Design (Sec.~\ref{sec:design})}} \\
\midrule
$N$                  & ZC sequence length \\
$r$                  & ZC root index (coprime to $N$) \\
$\mathbf{s}_r$       & Length-$N$ ZC sequence with root $r$ \\
$\mathbf{C}_r$       & ZC shift matrix $[\mathbf{s}_r,\mathbf{\Pi}\mathbf{s}_r,\dots]$ \\
$\mathbf{x}$         & Coefficient vector (payload mapped to shift domain) \\
$\mathbf{u}$         & Transmitted time-domain block (before CP) \\
$\mathbf{z}$         & ZC-shift-domain spectrum \\
$\mathbf{h}$         & Sampled CIR vector \\
$\tau$               & Equidistant shift spacing \\
$K = \lfloor N/\tau\rfloor$ & Number of active shift dimensions \\
$q_\ell = \ell\tau$  & Active shift index ($\ell=0,\dots,K-1$) \\
$a[\ell]$            & Unit-modulus coefficient at active dimension $\ell$ \\
$c[\ell]$            & Differential symbol at dimension $\ell$ \\
$\beta[d]$           & Folded path-profile energy at offset $d$ \\
$\widehat{\mathcal{D}}$ & Set of significant path offsets \\
$\Lambda_m[\ell]$    & Multipath-combined metric for candidate $m$ \\
$w_d$                & Path weight for offset $d$ \\
\midrule
\multicolumn{2}{c}{\textbf{LLC Design (Sec.~\ref{sec:amc})}} \\
\midrule
$\mathcal{U}$        & Set of registered users $\{1,\dots,K\}$ \\
$t_r$                & Start time of beacon $r$ \\
$T_b$                & Beacon airtime \\
$\tau_u$             & Single-trip propagation delay to user $u$ \\
$G_{\mathrm{pre}}$   & RTT-inclusive pre-budget \\
$W$                  & Continuous-backoff window length \\
$T_m$                & Uplink packet airtime under mode $m$ \\
$T_{\mathrm{SF}}$    & Superframe duration \\
$X_{u,r}$            & Random backoff drawn by user $u$ in round $r$ \\
$a_{u,r}, b_{u,r}$   & Packet arrival start/end at the gateway \\
$m_r$                & Common modulation scheme for superframe $r$ \\
$p_{u,r}$            & Power-control command for user $u$ in round $r$ \\
$\bar{p}_{u,r}$      & Exponentially weighted (EWMA) power state \\
$q_{u,r}$            & Power-control command (\texttt{UP/DOWN/HOLD}) \\
$\xi_u$              & CIR similarity (magnitude of normalized inner product) \\
$\mathcal{I}_{u,\max}[r]$ & Set of interferers overlapping user $u$ in round $r$ \\
$\widehat{\gamma}_{\mathrm{eff},u}[r]$ & Worst-case effective SINR \\
$\mathrm{PER}(m,\gamma,\xi)$ & Packet-error rate LUT \\
$m_r^\star$          & Candidate common MS for round $r+1$ \\
\bottomrule
\end{tabular}
\end{table}
}

\clearpage
\bibliographystyle{IEEEtran}
\bibliography{ref}

\end{document}